\documentclass[11pt,a4paper]{article}
\usepackage{amsmath}
\usepackage{amsfonts}
\usepackage{amssymb}
\usepackage{amsthm}
\usepackage{mathrsfs}
\usepackage{fullpage}
\usepackage{parskip}
\usepackage{dsfont}
\usepackage{graphicx}
\usepackage{subcaption}
\usepackage{comment}

\newtheorem{theorem}{Theorem}[section] 
\newtheorem{prop}{Proposition}[section]
\newtheorem{cor}{Corollary}[section]

\theoremstyle{definition}
\newtheorem{rem}{Remark}[section]

\newcommand{\R}{{\mathbb R}}

\newcommand{\Pp}{{\mathbb P}}
\newcommand{\E}{{\mathbb E}}
\newcommand{\V}{\mathrm {Var}}

\renewcommand{\theenumi}{\arabic{enumi}} 

\numberwithin{equation}{section}
\usepackage[round]{natbib}   
\begin{document}
\title{High-dimensional statistical arbitrage with factor models and stochastic control}
\author{Jorge Guijarro-Ordonez\footnote{Department of Mathematics, Stanford University. Email address: jguiord@stanford.edu.}
}
\date{October 2019}
\maketitle

\begin{abstract}
The present paper provides a study of high-dimensional statistical arbitrage that combines factor models with the tools from stochastic control, obtaining closed-form optimal strategies which are both interpretable and computationally implementable in a high-dimensional setting. Our setup is based on a general statistically-constructed factor model with mean-reverting residuals, in which we show how to construct analytically market-neutral portfolios and we analyze the problem of investing optimally in continuous time and finite horizon under exponential and mean-variance utilities. We also extend our model to incorporate constraints on the investor's portfolio like dollar-neutrality and market frictions in the form of temporary quadratic transaction costs, provide extensive Monte Carlo simulations of the previous strategies with 100 assets, and describe further possible extensions of our work. 
\end{abstract} 


\textbf{Word count:} 10,318 words.

\newpage
\section{Introduction}
Modeling of pairs trading based on stochastic control has been an active research topic in mathematical finance for the last few years. After the papers by \cite{jurek-yang} and \cite{orig-stoch-opt}, an increasing number of models have been proposed in this framework (see, for example, \cite{Chiu}, \cite{stochopt}, \cite{Timmermann}), in which generally they assume that some statistically-designed relation between the prices of two assets is a mean-reverting stochastic process and find a dynamic optimal allocation in continuous time in some version of the classical Merton framework. More recently, a number of papers have also studied the optimal entry and exit points when trading a couple of cointegrated assets, such as \cite{leung-stoploss}, \cite{LeiXu}, \cite{NgoPham}, and \cite{integral}. 

In the high-dimensional case, however, relatively little model-based research has been conducted. \cite{Cartea2016} and \cite{Tourin2016} investigate a multidimensional generalization of the model in \cite{stochopt} and apply stochastic control to solve a Merton-like problem in continuous time on a collection of cointegrated assets, with exponential utility and finite horizon. In a different direction which is not exactly statistical arbitrage, \cite{Cartea2018} addresses an optimal execution problem on a basket of multiple cointegrated assets, which they also solve with control techniques. Finally, without using stochastic control, the paper by \cite{avelee} carries out a data-based study of statistical arbitrage in the US equity market by proposing a factor model with mean-reverting residuals and a threshold-based strategy. This model is further analyzed and extended by \cite{papayeo}, who discuss risk control and develop an optimization method to allocate the investments given the trading signals.

The previous papers in this high-dimensional framework thus either apply stochastic control to a given mean-reverting process or use a factor model to construct this process and then choose the trading signals based on threshold rules, but none of them considers the combination of these two techniques. The present paper aims to fill this gap by providing a study of statistical arbitrage in a high-dimensional setting that combines factor models and the tools from stochastic control, considers transaction costs and statistical arbitrage constraints, and obtains closed-form optimal strategies which are interpretable and  easy to implement computationally. 

More precisely, in our framework an investor observes the returns of a high-dimensional collection of risky assets and, similar to \cite{avelee} and \cite{papayeo}, uses historical data to statistically construct  a factor model such that the cumulative residuals are assumed to be mean-reverting and following an Ornstein-Uhlenbeck process. However, unlike these previous studies, these residuals may be correlated and interdependent and, based on their behavior, the investor must decide how to optimally allocate her wealth in the risky assets and a riskless security so that the expected utility of her terminal wealth is maximized and she is market-neutral\footnote{In this paper we use the expression ``market-neutral'' as in \cite{avelee} to refer to factor neutrality.}. There are three main results in this paper:

First, for a big class of  statistically-constructed factor models that includes PCA we show how the investor may theoretically construct  market-neutral portfolios without solving any optimization problem (unlike the approach followed in \cite{papayeo} or \cite{boyd-mpo}, for example) provided that the factor model holds, and we show how this makes the optimal allocation problem analytically tractable and guarantees market-neutrality by construction. These portfolios are explicitly computable and depend quadratically on the factor model loadings and, to the best of our knowledge, using this construction to connect factor models and stochastic control theory is new. 

Second, using these explicit market-neutral portfolios as control variables, we show how the investor should trade optimally in continuous time to maximize either an exponential utility or a mean-variance objective, obtaining explicit analytic forms of the optimal strategies in both cases in this high-dimensional setting. The structure of these optimal strategies is related to the classical solution of the Merton problem and is affine in the deviation of the residuals from their statistical mean, thus giving a precise estimate of how much we should buy when the assets are underpriced and how much we should sell when they are overpriced, as in classical pairs trading. The coefficients are given by the solution of matrix Riccati differential equations and depend quadratically on the factor model loadings, and the strategies in both the exponential and the mean-variance case are surprisingly similar except for a non-myopic correction term that does not appear in the classical framework under a geometric Brownian motion. This arises from the fact that in our case the drift of the underlying Ornstein-Uhlenbeck process is stochastic. 

The structure and the techniques to find these affine strategies are thus similar in spirit to those in the affine process literature in finance (see \cite{DuffieAffine} for a broad survey),  to the more recent affine control literature in algorithmic trading (see, for example, \cite{CarteaBook} and the references therein), and to the literature on extensions of the Merton problem (see, for example, \cite{BenthKarlsen}, \cite{Liang}, \cite{Fouque} and \cite{Moutari}, which deal with a single risky asset in the context of the Schwarz model or in geometric Brownian motion with stochastic drift or volatility; and \cite{Brendel} and \cite{Bismuth}, which consider the multiasset case in the setting of geometric Brownian motion with uncertain drift). While the techniques that we use to find the optimal strategies are therefore classical, the framework and the results are new because the mean-reverting behavior of the underlying stochastic process arises from the residuals of a factor model and in the context of statistical arbitrage, and we consider the general case of an arbitrary number of assets with a market-neutrality restriction and a general matrix Ornstein-Uhlenbeck process. Moreover, the explicit solutions allow us to understand the dependence of the optimal strategies on specific elements of a statistical arbitrage strategy (such as the factor model, its loadings matrix and its connection with market-neutrality, and the mean-reversion speed of the residuals and their correlation structure), and to compare arbitrageurs with exponential and mean-variance utilities.    

Third, we extend the previous results in two directions by discussing how to incorporate into the model soft constraints frequently imposed by arbitrageurs such as dollar-neutrality, and also market frictions in the form of quadratic transaction costs, inspired by \cite{GarleanuPedersen2013} and \cite{GarleanuPedersen2016} and also by the more general quadratic transaction cost and linear price impact literature in portfolio theory (see, for example, \cite{Moreau} and \cite{linearquadratic} for some new research directions, and \cite{ObizhaevaWang2013}, \cite{Rogers}, \cite{AlmgrenChriss} and \cite{BertsimasLo} for some classical papers). In both extensions, we again find explicit analytic strategies which are easily interpretable, and which quantitatively correspond to quadratic corrections in the structure of the original optimal strategies (when adding soft constraints like dollar neutrality) or to ``tracking'' averages of the future original optimal portfolios (when adding quadratic transactions costs). Moreover, in both cases these new strategies depend quadratically on the loadings of the factor model. Again, the novelty of the results comes from the study of these questions (dollar neutrality, transaction costs, etc.) in a new context in which they are crucial (statistical arbitrage with an arbitrary number of assets following a general matrix Ornstein-Uhlenbeck process and a market-neutrality restriction, in particular using control techniques and a factor model), and this framework and the strategies that we find are new to the best of our knowledge.  

To conclude the paper with a more empirical analysis, we also perform extensive numerical simulations with a high-dimensional number of assets. This gives further insights about the  behavior of the previous strategies that are not obvious when looking at the corresponding equations, and allows us to understand the sensitivity of the model parameters and the dependence on the underlying factor model. This high-dimensional numerical study is also new with respect to the existing literature, and the main conclusions are that (1) the exponential-utility strategies are more profitable than the mean-variance strategies and they also take more extreme positions, (2) after some initial up and downs the sample paths of the different wealth processes progressively stabilize due to the asymptotic properties of the Ornstein-Uhlenbeck process, (3) increasing the risk-control parameters consistently produces a concentration of the distribution of the terminal wealth around smaller values, and (4) imposing market neutrality when the loadings of the factor model get bigger leads to more aggressive strategies whose terminal wealth has a higher variance.

The remainder of the article is organized as follows.  In section 2 we introduce our model, construct the market-neutral portfolios that make the problem analytically tractable, and formulate the control problems. Next, in section 3 we present the basic results under the exponential and the mean-variance frameworks, whereas section 4 extends these results by considering the addition of soft constraints and of quadratic transaction costs. Section 5 contains Monte Carlo simulations that provide further insight about the qualitative behavior of the found strategies, and section 6 presents the main conclusions and proposes future new directions of research. Finally, an appendix contains all the  proofs.  

\section{The model}
\subsection{Set-up and assumptions}

In the remainder of this paper we will consider the following general framework. We will assume that an investor observes the returns of a large number $N$ of risky assets and, like in classical portfolio theory based on stochastic control, she must decide how to dynamically allocate her wealth by investing in them or in a riskless asset with constant interest rate $r$ so that the expected utility of her wealth at a finite terminal time $T$ is maximized. However, unlike the classical framework and the existing literature, to do so she will execute a statistical arbitrage strategy based on a factor model, in which instead of trading depending on the state of the original returns she will trade depending on the behavior of the residuals, which will be the trading signals. For example, in the case of two assets, this is equivalent to classical pairs trading, in which the investor may perform a simple linear regression on the returns of two historically correlated securities and, depending on how far the oscillation of the residual is from its historical average, she decides if there is a mispricing and opens and closes long and short positions in the original assets in a market-neutral way. In this paper, we will study the generalization of this to the high-dimensional case of an arbitrary number of assets, in which we substitute the simple linear regression by a statistical factor model and we study the optimal allocations under the framework of stochastic control, assuming a mean-reverting stochastic model for the behavior of the residuals. 

More precisely, we make the following three general assumptions on how the investor will generate these residuals and what dynamics they will have:  

\begin{enumerate}
\item \textbf{Assumption 1:} The investor has computed a factor model for the returns of the risky assets, which will hold during the investment finite horizon and is given in differential form by
\begin{equation}\label{factormodel}
dR_t = \Lambda dF_t + dX_t,
\end{equation} 
where $R_t$ is the cumulative asset return process, $\Lambda$ is the (constant-in-time\footnote{We make this assumption for analytic tractability in our control framework, and given that the trading frequency is in general higher than the frequency at which these loadings will change significantly.}) loadings matrix, $F_t$ is the cumulative factor returns process, and $X_t$ is the cumulative residual returns process\footnote{Here we have written the factor model in a somewhat unusual differential form in terms of the cumulative residuals and returns because of notational simplicity for this section of the paper. In practice, however, the factor model will be estimated in discrete time, by replacing the differentials by the corresponding discrete increments (so, for instance, $dR_t$ should be replaced by the daily, hourly, etc. asset returns, $dF_t$ would be just the corresponding daily, hourly, etc. factors returns, and so forth). In any case we will only use this notation and framework in this section of the paper, and the reader may look at \cite{avelee} for essentially the same continuous/discrete time framework and some estimation techniques. }.

\item \textbf{Assumption 2:} This factor model has been computed statistically by using some version of PCA\footnote{See \cite{PelgerRPPCA} for some new versions of PCA that might be particularly interesting for this problem.}, so the rows of $\Lambda$ are the largest eigenvectors of some square matrix and the discrete-time version of $dF_t$ (i.e., the daily, hourly, etc. factors returns) is  computed by linearly regressing the discrete-time version of $dR_t$ (i.e., the daily, hourly, etc. assets returns) on some rescaling of $\Lambda$, so 
\begin{equation}\label{factorestimation}
dF_t = \tilde{\Lambda}dR_t
\end{equation}
for some matrix $\tilde{\Lambda}$. However, the only fact we will need about this assumption is that (\ref{factorestimation}) holds for some matrix $\tilde{\Lambda}$, which allows for a bigger class of factor models than classical PCA.

\item \textbf{Assumption 3:} The process $X_t$ given by the cumulative residuals is mean-reverting. In particular, for analytic tractability we assume that it is a matrix $N$-dimensional Ornstein-Uhlenbeck process satisfying the following stochastic differential equation with known parameters
$$ dX_t = A(\mu - X_t)dt+\sigma dB_t,$$
where $A$ is a constant $N$-dimensional square matrix, $\mu$ is a constant $N$-dimensional vector, $B_t$ is a vector of $m$ independent Brownian motions in the usual complete filtered probability space $(\Omega, \mathcal{F},\Pp,(\mathcal{F}_t)_{0\leq t\leq T})$, and $\sigma$ is a constant $N\times m$ matrix such that the instantaneous covariance matrix $\sigma\sigma'$ is invertible.
\end{enumerate}

The previous framework thus combines high-dimensional statistical arbitrage, factor models and stochastic control in a way which is new to the best of our knowledge, and it extends several models in the existing literature. For example, statistical arbitrage models based on a more particular case of Assumptions 1, 2, 3 (in which the residuals are assumed to be independent one-dimensional Ornstein-Uhlenbeck processes, so $A$ and $\sigma$ are diagonal) and in which no stochastic control methods are applied have been studied empirically  in the US equity market by \cite{avelee} and \cite{papayeo}. In a different direction, if we consider the particular case of removing the factor model by making $\Lambda = 0$, we have the situation in which the returns are globally mean-reverting following a matrix Ornstein-Uhlenbeck process, which has also been studied empirically and analytically using stochastic control techniques in the context of optimal execution in \cite{Cartea2018}.

\subsection{Making the model tractable and imposing market-neutrality}

Unlike the classical literature on portfolio choice based on stochastic control, choosing as control variables the amount of capital that the agent invests in each of the $N$ risky assets of the previous framework might make the optimal allocation problem analytically intractable. Indeed, since we only have information about the dynamics of the residuals and not directly about the returns like in the classical framework, these residuals are not independent, and the factors depend on the returns, the classical approach would lead to complicated interdependencies. Moreover, since the investor is executing a statistical arbitrage strategy, we would need to incorporate additional market neutrality constraints\footnote{Here we use market-neutrality in the sense of \cite{avelee}.} so that the returns of the strategy do not depend on the model factors, but just on the idiosyncratic component of the model given by the residuals. This would complicate the problem further, and would require numerical optimization methods as done in \cite{papayeo}.

In this paper, on the contrary, we approach both problems simultaneously and we solve them analytically by following a new approach. This is based on the following proposition, which shows that, by using the original $N$ risky assets, it is possible to construct analytically $N$ market-neutral portfolios whose returns only depend on one coordinate of $X$, which simplifies the complexity of the problem and makes it analytically tractable:\\

\begin{prop}\label{prop:marketneutrality}
Under the previous assumptions, it is possible to construct explicitly $N$ market-neutral portfolios such that investing any real number $\pi_{it}$ of dollars in the $i$-th one at time $t$ yields an instantaneous return of $\pi_{it}dX_{ti}$ (and hence a combined return of $\pi_{t}\cdot dX_{t}$).

Moreover, the total amount of capital invested at time $t$ by doing so is $\pi_t\cdot p$ for an explicit constant-in-time vector $p\in\R^N$, which depends quadratically on the factor model loadings.
\end{prop}
\begin{proof}
The mathematical construction of the market-neutral portfolios  under the given assumptions is surprisingly straightforward and involves just a linear projection. Indeed, (\ref{factormodel}) implies that
$$dR_{ti} = \sum_j\Lambda_{ij}dF_{tj}+dX_{ti},$$
whereas (\ref{factorestimation}) yields 
$$dF_{tj} = \sum_k\tilde{\Lambda}_{jk}dR_{tk}.$$
Combining the two previous equations we find that, for $c_{ik}:=\sum_j\tilde{\Lambda}_{jk}\Lambda_{ij}$,
$$dR_{ti} = \sum_k\left(\sum_j\tilde{\Lambda}_{jk}\Lambda_{ij}\right)dR_{tk}+dX_{ti}=\sum_kc_{ik}dR_{tk}+dX_{ti}.$$
Thus, if at time $t$ we hold the (explicitly constructible) constant-in-time portfolio given by 
$$\tilde{p}_i:=(-c_{i1},-c_{i2},\ldots,-c_{i,i-1},1-c_{ii},-c_{i,i+1},\ldots, -c_{iN})$$
(i.e., we invest $-c_{i1}$ dollars in the first asset, $-c_{i2}$ dollars in the second one, and so on), we automatically obtain an instantaneous return of $dX_{ti}$, which is market neutral and depends only on the $i$th coordinate of the process $X_t$. Further, from the above equations it is also obvious that for any real number $\pi_{it}$, $ \pi_{it}\tilde{p}_i$ will also be market-neutral and yielding a return of $\pi_{it}dX_{ti}$, and the same applies to $\sum_i\pi_{it}\tilde{p}_i$, which will have a return of $\sum_i\pi_{it}dX_{ti}=\pi_t\cdot dX_t$.

Finally, regarding the last part of the statement just observe that the total amount of capital invested in the strategy $\pi_t = (\pi_{it})_{1\leq i\leq N}$ at time $t$ is simply $$\sum_i(\pi_{it}\tilde{p}_i)\cdot \mathds{1} = \sum_i\pi_{it}(\tilde{p}_i\cdot \mathds{1})=\pi_t\cdot p$$
where $p:=(\tilde{p}_i\cdot \mathds{1})_{1\leq i\leq N}$, which concludes our proof.
\end{proof}

\begin{rem}
Note in particular that, if $\Lambda$ or $\tilde{\Lambda}$ are sparse matrices, then most of the $c_{ik}$ in the above construction will be 0, so the investor will be investing in a few number of assets in each market-neutral portfolio and this could significantly reduce his transaction costs while rebalancing his positions. In particular, \cite{PelgerSparse} discusses a way of obtaining this kind of sparse factor model.
\end{rem}

The key consequence of the above proposition is that, if we choose as control variables the amount of capital $\pi_t$ that we wish to invest in  these $N$ market-neutral portfolios (instead of directly in the original assets) at time $t$, the dynamics of the problem get simpler, they only depend separately on the coordinates of $X$, and we have market-neutrality by construction. This solves simultaneously the two problems we discussed before and allows us to connect stochastic control and the factor model in a simple way, and it is therefore the approach which we will adopt in the remainder of this paper.

Note also that, under these new control variables, all the information about the factor model and in particular about its loadings matrix is now encoded in the vector $p$, which will play an important role in the remaining sections. Moreover, some statements about the strategies must be rewritten in terms of it within this new framework. For instance, in the new setting a strategy $(\pi_t)_{0\leq t\leq T}$ is dollar-neutral at $t$ if $p\cdot\pi_t=0$, since as we mentioned before $p\cdot\pi_t$ is the total capital spent at time $t$.

\subsection{Formulation of the control problems}

Under the previous framework, now we formulate rigorously the control problems we will study in the paper. We suppose that the investor executes the following trading strategy: at each time $t\in[0,T]$, she invests $\pi_t$ dollars in the risky market-neutral portfolios we constructed in Proposition \ref{prop:marketneutrality}, and she invests her remaining wealth in the risk-free asset with constant interest rate $r$, so that the resulting strategy is self-financing. Thus, assuming for the moment no market frictions or  constraints (which will be both considered in section 4), the evolution of her wealth is given by the equation
\begin{equation}\label{Wealth}
dW_t =\pi_t\cdot dX_t+(W_t-\pi_t\cdot p)rdt
\end{equation}
and she aims to choose $\pi_t$ to maximize the expected utility of her terminal wealth, which is given by $u(W_T)$ for a fixed utility function $u$. 

Supposing further that she trades continuously in time, this means that mathematically she must solve the high-dimensional non-linear stochastic optimization problem given by
\begin{equation}\label{ProblemCont}
H(t,x,w)=\sup_{\pi\in\mathcal{A}_{[t,T]}}\E_{t,x,w}\left[u(W_T)\right]
\end{equation}
subject to 
$$dW_t = \left(\pi_t'A(\mu - X_t)+(W_t-\pi_t'p)r\right)dt+\pi_t'\sigma dB_t$$
$$dX_t = A(\mu - X_t)dt+\sigma dB_t,$$
where  $'$ indicates transposition, and the admissible set $\mathcal{A}_{[t,T]}$ is the set of all the $\mathcal{F}_s$-predictable and adapted processes $(\pi_s)_{s\in[t,T]}$ in $\R^N$ with the minimal technical restrictions that $\E[\int_t^T||\pi_s||^2ds]<\infty$ (so Ito's formula may be applied and doubling strategies are excluded) and the above SDEs have a unique strong solution.

Finally, the associated dynamic programming equation of the problem is non-linear and $(N+2)$-dimensional, and is given by
\begin{multline}\label{DPE}
0=\partial_tH + (\mu - x)'A'\nabla_xH+\frac{1}{2}\mathrm{Tr}(\sigma\sigma'\nabla_{xx}H)\ +\\ 
\sup_\pi\left(\left(\pi'A(\mu - x)+(w-\pi'p)r\right)\partial_wH+\frac{1}{2}\pi'\sigma\sigma'\pi\partial_{ww}H + \pi'\sigma\sigma'\nabla_{xw}H \right)
\end{multline}
with terminal condition $H(T,x,w)=u(w)$.

The problem is therefore formally related to the classical Merton framework, but instead of a geometric Brownian motion there is a multidimensional Ornstein-Uhlenbeck process which makes it impossible to combine the dynamics of $W$ and $X$ into a single equation and to get rid of the $N$-dimensional state variable $x$. Moreover, unlike the previous studies on extensions of the Merton problem with an Ornstein-Uhlenbeck process discussed in section 1, in (\ref{ProblemCont}) and (\ref{DPE}) the mean-reverting behavior of the underlying stochastic process arises in the context of statistical arbitrage and from the residuals of a factor model (which is encoded in the vector $p$ of the equations above and which will play an important role in the following sections), and we consider the general case of an arbitrary number of assets with a market neutral restriction. Furthermore, the model will be extended in section 4 to incorporate other important features of statistical arbitrage strategies, like dollar neutrality restrictions and transaction costs, and we will analyze the impact of the factor model on these extensions.

Quite surprisingly, the previous problems admit  interpretable closed-form solutions -- which is computationally useful in this high-dimensional setting, and which allows us to understand the influence of the model parameters and especially of the factor model -- in the cases in which the utility is exponential or of a Markowitz-inspired mean-variance type, but not for other usual choices of utility functions, like the HARA family. We show this in the following two sections, first for the simple setup of (\ref{ProblemCont}) and (\ref{DPE}) in section 3, and then extending the model in section 4 to incorporate soft constraints on the investor's portfolio and quadratic transaction costs.

\section{The frictionless results}
In this section we present the closed-form, optimal strategies for the  problem given by (\ref{ProblemCont}) and (\ref{DPE}) in the cases in which the utility is exponential or of a mean-variance type, discussing the former in the first subsection and the latter in the second one. 

\subsection{The exponential utility case}

In the first setting, the explicit description of the optimal strategy is given by the following main theorem (see \cite{Cartea2016} and \cite{Tourin2016} for related results with an exponential utility):   \\

\begin{theorem}
Under an exponential utility (so $u(w)=-e^{-\gamma w}$ for some $\gamma>0$) and the technical condition described in our verification theorem (Proposition 3.2 below), the optimal portfolio to have at time $t$ according to (\ref{ProblemCont}) is explicitly computable and given by
 $$\pi^*_t = (\sigma\sigma')^{-1}\frac{A(\mu-X_t)-pr}{\gamma e^{r(T-t)}} +\frac{A'(\sigma\sigma')^{-1}}{\gamma e^{r(T-t)}}
\left((A(\mu-X_t)-pr)(T-t)-Apr\frac{(T-t)^2}{2}
\right).
$$\end{theorem}

The result follows from the following two propositions, whose proof is given in appendix A.1 using classical stochastic control techniques:\\

\begin{prop} [Solving the PDE] The value function $H$ associated to (\ref{ProblemCont}) and (\ref{DPE}) when $u(w)=-e^{-\gamma w}$ is explicitly computable and admits the probabilistic representation $H(t,x,w)=-\exp(-\gamma(we^{r(T-t)}+h(t,x)))$ where
$$h(t,x) = \E_{t,x}^*\left[\int_t^T \frac{1}{2\gamma}(A(\mu-Y_s)-pr)'(\sigma\sigma')^{-1}(A(\mu-Y_s)-pr)\ ds\right]$$
and $dY_t=rpdt+\sigma dB_t^*$ for a new Brownian motion $B^*$ under a new probability law $\Pp^*$. The associated optimal control in feedback form is then \begin{equation}\label{optimalpi}
\pi^* = -(\sigma\sigma')^{-1}\frac{\mathcal{D}H}{\partial_{ww}H}
\end{equation} where $\mathcal{D}H=\left(A(\mu - x)-pr\right)\partial_wH + \sigma\sigma'\nabla_{xw}H$.   \\
 \end{prop}
 
\begin{prop}[Verification]  The strategy given in Theorem 3.1. is indeed optimal if 
$$4\max_{0\leq s\leq T}||\Lambda_0(s)||<1\quad\text{and}\quad 32\max_{0\leq s\leq T}||\Lambda_1(s)||<1 ,$$
where $\Lambda_0(s)$ and $\Lambda_1(s)$ are the diagonal matrices containing, respectively, the eigenvalues of $\Omega^{1/2}(C_0+C_0')\Omega^{1/2}(s)$ and $\Omega^{1/2}C_1C_1'\Omega^{1/2}(s)$, for
$$C_0(s)=A'(\sigma\sigma')^{-1}A(I_N+A(T-s)),\quad C_1(s)=A'(\sigma\sigma')^{-1}(I_N+A(T-s))\sigma $$
$$\Omega(s)= \int_0^{s} e^{-A(s-u)}\sigma\sigma' e^{-A'(s-u)}du.$$
\end{prop}

Besides being a closed-form  strategy which is easily implementable in our high-dimensional setting,  the above optimal portfolio is also interpretable. Indeed, the first term of the optimal policy is Merton-like since it represents the drift of the underlying process (which here is stochastic unlike in the classical geometric Brownian motion) minus the adjusted risk-free rate (which here depends on the loadings of the factor model via $p$). This is divided by a measure of the volatility (which is given by $\sigma\sigma'$, the instantaneous quadratic covariation of $X$) and the Arrow-Pratt coefficient of absolute risk-aversion of the value function with respect to the wealth $w$ (i.e., $-\partial_{ww}H/\partial_{w}H$), which is the product $\gamma e^{r(T-t)}$, where $\gamma$ is the risk aversion parameter of the utility function and the factor $e^{r(T-t)}$ measures the gains from interest between $t$ and $T$. 

On the other hand, the second summand is a non-myopic correction term which again depends linearly on the drift of $X$, and whose effect vanishes when we approach the terminal time $T$. Moreover, while the first term does not depend explicitly on the terminal time $T$, this correction term does, reflecting the fact that, since there are non-zero interest rates and moreover the behavior of the residuals is oscillating, the investor must keep in mind the final horizon to decide if she bets on the mean-reversion cycle before that time. Finally, as the risk-aversion parameter $\gamma$, the instantaneous volatility $\sigma\sigma'$, or the interest rate $r$ increase, the optimal portfolio vector $\pi_t^*$ gets closer to 0, implying that the investor will simply invest most of her wealth in the riskless asset.

The above strategy is also intuitive within our framework of statistical arbitrage with a factor model, and sheds further light on the problem. Indeed, note that the current state of the residual process $X_t$ only appears in the strategy through the terms in $A(\mu-X_t)$, which essentially tells us to invest more in the risky assets the further their residuals are from their historical mean $\mu$ and in a way proportional to the historical mean reversion speed given by $A$, like in classical pairs trading. Moreover, all the remaining terms depend jointly on the  factor model and the interest rate through the term $pr$, which reflects the cost of the leverage associated to imposing market-neutrality through the loadings of the factor model. In particular, note that, the bigger the loadings of the factor model are and hence the bigger $p$ is, the more the agent will need to invest to achieve market neutrality (again like in pairs trading with a big beta) and the bigger her leverage will be, and this will affect the optimal strategy depending on the interest rate $r$.

Finally, regarding the technical optimality conditions, intuitively they arise from the fact that $H(t,X_t,W_t^*)$, the value function evaluated at the wealth process $W_t^*$ corresponding to the optimal strategy, may blow up because of the exponential function coming from the exponential utility. In particular, since $W_t^*$ ends up being an Ito process depending quadratically on $X_t$ and $X_t$ is Gaussian, the term $\exp(-\gamma W_t^*e^{r(T-t)})$ is related to the moment generating function of a chi-squared distribution, which blows up far away from 0. Thus, these technical conditions are just ensuring that the corresponding functions are integrable. Interestingly, this does not depend on the  risk-aversion parameter $\gamma$, the interest rate $r$, or the factor model used (captured by $p$), but just on the parameters of $X$ and the terminal time $T$.

\subsection{The mean-variance case}

In the second, Markowitz-inspired mean-variance framework, the investor tries to maximize her expected terminal wealth, but at the same time she continuously penalizes at each instant the instantaneous variance of her wealth process according to a time-dependent volatility-aversion function $\gamma(t)$. The optimal strategy in this case is again available in closed form and interpretable and, for an appropriate choice of this volatility-aversion function, we obtain the same optimal policy as in the exponential case but without the correction term. This is shown in the following theorem, whose proof is given in appendix A.2:\\

\begin{theorem}
If $\gamma(t)$ is continuous and positive on $[0,T]$, the problem in (\ref{ProblemCont}) with the following mean-variance objective function 
$$H(t,x,w)=\sup_{\pi\in\mathcal{A}_{t,T}}\E_{t,x,w}\left[ W_T-\int_t^T\frac{\gamma(s)}{2}\frac{d}{d\tau}\left.\V_s(W_\tau)\right|_{\tau=s} ds\right]
$$
has explicit optimal portfolio at $t$ given by 
$$\pi^*_t = (\gamma(t)\sigma\sigma')^{-1}\left(A(\mu-X_t)-pr\right)e^{r(T-t)}.$$

In particular, for $\gamma(t)=\gamma e^{2r(T-t)}$, the above optimal policy is the same as the first term of the corresponding one in Theorem 3.1. 
\end{theorem}


Regarding the interpretation of the mean-variance strategy within our context of statistical arbitrage and its connection with the exponential-utility arbitrageur, there are two important remarks. 

First, as we mentioned, the new optimal strategy is the same as the myopic part of the exponential case modulo the value of $\gamma(t)$. In particular, this means that, unlike the exponential arbitrageur, the mean-variance arbitrageur will not take into account the expected number of mean-reversion cycles until the terminal time $T$. Moreover, for a non-zero interest rate and a constant 
volatility aversion $\gamma(t)$, the mean-variance arbitrageur is more aggressive than the corresponding exponential investor with the same $\gamma$, since she will invest significantly more capital (quantitatively, by a factor of $e^{2r(T-t)}$) in going long or short, taking more aggresive positions the higher the interest rate is and the sooner it is with respect to the terminal date.

Second, the optimal strategy has two components like in section 3.1: one term in $A(\mu-X_t)$ which measures how far the residuals are from their historical mean and how fast they will mean-revert (like in classical pairs trading), and a second term in $pr$ linked to the factor model, which measures the cost of the leverage associated to imposing market neutrality. In particular, note that, the bigger the loadings of the factor model are (and hence the bigger $p$ is), the more aggressive the positions will be and the more leverage the investor will have if $r\neq 0$.  

\section{Two extensions}

In this section of the paper, we complete the picture described in the previous two sections by considering two important extensions within the context of statistical arbitrage with a factor model. In the first subsection, we show how to incorporate in the above strategies soft constraints frequently imposed by arbitrageurs with the example of dollar-neutrality. In the second one, we introduce market frictions in the form of quadratic transaction costs. In both cases, we obtain again  closed-form analytic solutions which are interpretable, convenient from a computational perspective in our high-dimensional setting, and which shed further light on the influence of the factor model and its connection with market neutrality.

\subsection{Incorporating soft constraints on the admissible portfolios}

While imposing restrictions on the portfolios by introducing hard constraints directly on the admissible set $\mathcal{A}_{t,T}$ leads in general to control problems that must be solved numerically (and hence potentially unfeasible in a high-dimensional setting since in applications the number of assets ranges from hundreds to thousands), it is still possible to impose many additional soft constraints in the two frameworks of section 3 without significantly increasing the difficulty of the problems, by just adding a carefully chosen penalty term to the corresponding objective function.

As an illustration of this, we give in the next corollary the corresponding optimal strategies when a dollar-neutrality restriction is softly enforced. To do so, recall that, within the framework of section 2 that imposed market-neutrality within the factor model, a strategy $\pi_t$ is dollar neutral if $p\cdot\pi_t=0$, which means that the total amount of capital invested at time $t$ is 0. Hence, we can softly enforce dollar neutrality by replacing the wealth process of Theorems 3.1 and 3.2 by the penalized wealth process defined by $d\tilde{W}_t:=dW_t-\alpha(t)(\pi_t\cdot p)^2/2dt$ for a certain general time-dependent penalty function $\alpha(t)$. This penalizes non dollar-neutrality (i.e., $\pi_t\cdot p\neq 0$) at each time and is quadratic to simplify the optimization process.

The proof follows the same lines as in the previous two cases and is obtained from them by small modifications, so we will omit it for the sake of brevity.\\

\begin{cor} Suppose that dollar neutrality is softly enforced  by replacing the wealth process of Theorems 2.3.1 and 2.3.2 by the penalized wealth process defined by $d\tilde{W}_t:=dW_t-\alpha(t)(\pi_t\cdot p)^2/2dt$. Then
\begin{enumerate}
\item The problem with mean-variance utility has optimal portfolio at $t$ given by
$$\pi^*_t = (\gamma(t)\sigma\sigma'+\alpha(t) pp')^{-1}\left(A(\mu-X_t)-pr\right)e^{r(T-t)}.$$
\item The problem with exponential utility has optimal portfolio at $t$ given by
$$\pi^*_t = (\gamma e^{r(T-t)}\sigma\sigma'+\alpha(t) pp')^{-1}\left(A(\mu-X_t)-pr -\gamma\sigma\sigma'(b(t)+c(t)X_t)\right).
$$

where $c(t)$ is an $N\times N$ symmetric matrix and $b(t)$ is an $N$-dimensional vector, vanishing when $t\rightarrow T$, and with coordinates depending on $A,\sigma,rp,\gamma$ and $\alpha(t)$. In particular, $c(t)$ is given by the solution of the matrix Riccati ODE
$$0=\partial_tc-A'c-cA-\gamma c\sigma\sigma'c+e^{r(T-t)}(A+\gamma\sigma\sigma'c)'M(t)(A+\gamma\sigma\sigma'c)$$
and $b(t)$ is the solution of the linear system of ODEs 
$$ 0 = \partial_tb-A'b+cA\mu-e^{r(T-t)}(A+\gamma\sigma\sigma'c)M(t)(A\mu-pr-\gamma\sigma\sigma'b)-\gamma c\sigma\sigma'b,$$
both with terminal conditions $b(T)=c(T)=0$ and where $M(t)=(\gamma\sigma\sigma'e^{r(T-t)}+\alpha(t)pp')^{-1}$.
\end{enumerate}
\end{cor}

The resulting optimal policies have therefore the same structure as the two previous strategies of section 3, but now the additional term $\alpha(t) pp'$ has been introduced in the inverse to enforce the dollar-neutrality condition. This again depends on the factor model via $p$ and is related to how extreme the capital positions will be because of the market-neutrality restriction, which depends directly on the loadings matrix and hence on $p$. Note in particular that, the bigger the loadings of the factor model are, the bigger $\alpha(t) pp'$ will be and hence the bigger the impact of the dollar neutrality restriction will be. 

\subsection{Incorporating quadratic transaction costs}

In this subsection, we extend our model to incorporate market frictions in the form of transaction costs, which play a crucial role when executing statistical arbitrage strategies. We consider in particular quadratic transaction costs, which are in general a measure of price impact or illiquidity and which make the model anaytically tractable.

To do so, rather than looking at the amount of capital $\pi_t$ invested in the risky assets at time $t$ as the control variables, we consider the trading intensity $I_t$ at which these investements will be made at time $t$, which is given by $d\pi_t=I_tdt$. We can now adapt to our setting the model for transaction costs introduced in \cite{GarleanuPedersen2016}, who posit, providing market microstructural justification and referring to empirical research, that these transaction costs at time $t$ may be represented quadratically as $I_t'CI_t$ for a certain symmetric positive-definite matrix $C$\footnote{The assumption that $C$ is symmetric is without loss of generality, since if the transaction costs are given by $I_t'\tilde{C}I_t$ for a non-symmetric $\tilde{C}$, then one can see that by considering the symmetric part of $\tilde{C}$ (given by $C:=(\tilde{C}+\tilde{C}')/2$) we have that $I_t'\tilde{C}I_t=I_t'CI_t$.}, which essentially comes from the assumption that the price impact of the investor's actions is linear on its trading intensity $I_t$. 

Under this framework, we rewrite the performance criteria of Theorem 3.2 by incorporating the adverse effect caused by these transaction costs on the investor's wealth as a running penalty, obtaining the stochastic optimization problem given by
\begin{equation}\label{costscontinuous}
H(t,x,w,\pi)=\sup_{I\in\mathcal{A}^*}\E_{t,x,w,\pi}\left[ W_T-\int_t^T\frac{\gamma(s)}{2}\frac{d}{d\tau}\left.\V_s(W_\tau)\right|_{\tau=s} ds-\frac{1}{2}\int_t^TI_s'CI_sds\right].
\end{equation}
As we mentioned, in this new problem the control variable is $I$; $t,x,w,\pi$ are now state variables; and $\mathcal{A}^*$ is the set of all $\mathcal{F}$-adapted predictable processes $I_t$ such that the corresponding SDEs have a unique strong solution for any initial data and both $I_t$ and the resulting $\pi_t$ given by $d\pi_t=I_tdt$ are again in $L^2(\Omega\times[0,T])$. Thus, the investor aims to maximize her terminal wealth, but penalizing at each instant both for the risk of her strategy (measured by the instantaneous variance of her wealth process) and for the price impact caused by her actions (reflected in the quadratic transaction costs).

In this new setting, it is again possible to find explicitly the optimal strategy that the investor should follow, which is described in detail in the next theorem:\\

\begin{theorem}
If $\gamma(t)\geq 0$ and is continuous, the optimal strategy in the above problem ``tracks'' a moving aim portfolio $\mathrm{Aim}(t,X_t)$ with a tracking speed of $\mathrm{Rate}(t)$, according to the following ODE describing the evolution of the optimal trading intensity $I_t=d\pi_t/dt$
$$I_t=\mathrm{Aim}(t,X_t)+\mathrm{Rate}(t)\pi_t,$$  
where $\mathrm{Rate}(t)$ is a $N\times N$ negative-definite matrix tending to 0 when $t\rightarrow T$\footnote{and given by the solution of a matrix Riccati ODE specified in the Porposition 2.4.2 below.}, and $\mathrm{Aim}(t,X_t)$ admits the probabilistic representation
$$\mathrm{Aim}(t,x)=\int_t^Tf(s)\E_{t,x}[\mathrm{Frictionless}(s)]ds $$
where $\mathrm{Frictionless}(s)$ is the optimal portfolio at time $s$ in the frictionless case of section 3.2. and $f(s)$ is a certain averaging function given in Proposition 4.3 below.

Furthermore, the optimal portfolio is then
$$\pi_s^*=\pi_t+\int_t^s:\exp\left(\int_u^s\mathrm{Rate}(v)dv\right):\mathrm{Aim}(u,X_u)du,$$
where the notation $:\exp(\int_u^t\cdot ds):$ represents the time-ordered exponential\footnote{Recall that the time-ordered exponential of a time-dependent matrix $A(s)$ is defined as $:\exp(\int_u^tA(s)ds):=\lim_{||\mathcal{P}||\downarrow 0}\prod_{i=1}^{n_\mathcal{P}}\exp(A(t_i)\Delta t_i)$, where $\mathcal{P} := \{u=t_0,t_1,\ldots, t_n=t\}$ is a partition of $[u,t]$, $\Delta t_i:=t_i-t_{i-1}$, and the product is ordered increasingly in time. If $A(s)$ is a scalar, then obviously $:\exp(\int_u^tA(s)ds):=\exp(\int_u^tA(s)ds)$.}.\\
\end{theorem}

\begin{rem} If in particular the investor has constant volatility aversion (so $\gamma(t)=\gamma$), the matrix Riccati ODE is explicitly solvable and 
$$\mathrm{Rate}(t)=C^{-1/2}D\tanh(D(t-T))C^{1/2}$$
for $D:=(\gamma C^{-1/2}\sigma\sigma'C^{-1/2})^{1/2}$.
 Moreover, if the transaction costs are proportional to the volatility (i.e., $C=\lambda \sigma\sigma'$ for $\lambda>0$, see \cite{GarleanuPedersen2013,GarleanuPedersen2016} for a market microstructural justification) then this rate is indeed a scalar given by $\sqrt{\frac{\gamma}{\lambda}}\tanh\left(\sqrt{\frac{\gamma}{\lambda}}(t-T)\right)$ and $:\exp\left(\int_u^s\mathrm{Rate}(v)dv\right):=\cosh\left(\sqrt{\frac{\gamma}{\lambda}}(s-T)\right)/\cosh\left(\sqrt{\frac{\gamma}{\lambda}}(u-T)\right)$. 
\end{rem}
The result follows from the following sequence of three propositions, which are proved in appendix A.3:\\

\begin{prop}[Conjectured solution] The solution of the HJB equation associated to the problem is $H(t,x,w,\pi)=e^{r(T-t)}w+\frac{1}{2}\pi'a(t)\pi+\pi'b(t,x)+d(t,x)$ if there exist a $N\times N$ symmetric matrix $a(t)$, a $N$-dimensional vector $b(t,x)$ and a scalar function $d(t,x)$ satisfying
\begin{enumerate}
\item The matrix Riccati ODE
\begin{equation}\label{RiccatiCosts}
\partial_t a -\gamma(t)\sigma\sigma'+aC^{-1}a=0
\end{equation}  
with terminal condition $a(T)=0$.
\item The vector-valued and the scalar linear parabolic PDEs 
\begin{equation}\label{LPCosts1}
(\partial_t+\mathcal{L}_X)b+e^{r(T-t)}(A(\mu-x)-rp)+a'C^{-1}b=0
\end{equation}  
\begin{equation}\label{LPCosts2}(\partial_t+\mathcal{L}_X)d+\frac{1}{2}b'C^{-1}b =0
\end{equation}  
with terminal conditions  $b(T,x)=d(T,x)=0$ and where $\mathcal{L}_X$ is the infinitesimal generator of $X$, acting coordinatewise.
\end{enumerate}
The hypothesized optimal trading intensity at $(t,x,w,\pi)$ is then $I^*=C^{-1}(a(t)\pi+b(t,x))$.\\
\end{prop}

\begin{prop}[Existence of solutions].  

\begin{enumerate}
\item If the volatility aversion $\gamma(t)\geq 0$ and is continuous, then the Riccati equation (\ref{RiccatiCosts}) has a symmetric, bounded and negative definite solution on all $[0,T]$. In particular, for $\gamma(t)=\gamma$, the solution is 
$$a(t)=C^{1/2}D\tanh(D(t-T))C^{1/2}$$ for $D:=(\gamma C^{-1/2}\sigma\sigma'C^{-1/2})^{1/2}$.

\item Moreover, under this condition  the parabolic PDEs (\ref{LPCosts1}) and (\ref{LPCosts2}) have a unique solution satisfying a polynomial growth condition in $x$, and this solution admits the probabilistic representation 
\begin{equation}\label{Probrepb}
b(t,x)=\E_{t,x}\left[\int_t^T:\exp\left(\int_t^sa'(u)C^{-1}du\right):e^{r(T-s)}(A(\mu-X_s)-rp)ds\right]
\end{equation}
\begin{equation}\label{Probrepd}
d(t,x)=\frac{1}{2}\E_{t,x}\left[\int_t^Tb(t,X_s)'C^{-1}b(t,X_s)ds\right]
\end{equation}

Furthermore, $b$ has linear growth in $x$  and $d$ has quadratic growth in $x$, both uniformly in $t$.\\
\end{enumerate}
\end{prop}

\begin{prop}[Verification] Under the assumptions of the previous proposition, the trading intensity given in Theorem 4.1 is indeed optimal with the choices
$$\mathrm{Aim}(t,x)=C^{-1}b(t,x),\quad \mathrm{Rate}(t)=C^{-1}a(t),\quad f(s)=C^{-1}:\exp\left(\int_t^s\mathrm{Rate}(u)'du\right):\gamma(s)\sigma\sigma'.$$
\end{prop}

The interpretation of the above strategy, which is again explicit and hence easily implementable in practice, is intuitive and complementary to the infinite-horizon model of \cite{GarleanuPedersen2016}: in the presence of quadratic transactions costs, the investor trades with a certain decreasing rate $\mathrm{Rate}(t)$ towards an aim portfolio $\mathrm{Aim}(t,X_t)$ depending on the time and the mean-reversion state of the signals $X_t$. This aim portfolio is given by a weighted average of the future optimal strategies in the frictionless case, reflecting the fact that now trading is not free and thus to enter a trade the investor must weight the future outcomes derived from the strategy. Moreover, as shown in the above remark, the trading rate is bounded by 1 because of the properties of $\tanh$, depends on $t$ unlike the infinite-horizon case, and is naturally decreasing in the transaction cost parameter $\lambda$ (or in general in $C$) and increasing in the volatility aversion parameter $\gamma$.

Finally, regarding the influence of the factor model and the imposition of market neutrality in this new setting, note that $\mathrm{Rate}(t)$ is insensitive to it, since it only depends on the risk aversion parameter $\gamma$, the volatility of the residual process $\sigma\sigma'$, and the transaction cost matrix $C$. Similarly, in $\mathrm{Aim}(t,x)$ it only appears through the term $\E_{t,x}[\mathrm{Frictionless}(s)]$ and hence only when considering the future optimal strategies in the frictionless case, which has been described in section 3. Likewise, the residual process $X_s$ only affects the strategy through this same term and hence, as seen in section 3 when studying these frictionless cases, only through the distance between this residual and its historical mean, like in classical pairs trading.

\section{Monte Carlo simulations}
We conclude the paper by providing some high-dimensional numerical simulations that give new insights about the behavior of the previously discussed strategies and their sensitivity to the different parameters, emphasizing in a separate simulation the role of the factor model and its connection with market-neutrality. To this end, we  first simulate $X$ in the case of $N=100$ residuals by using exact Monte Carlo sampling along a discrete time grid, and we execute the previous strategies for some parametric choices of $X$ and some values of $p$ to compute sample paths of $\pi_t$ and $W_t$ and histograms of the terminal Profit \& Loss (P\&L). We have however opted to defer experiments with real data to a separate paper, since examining carefully the delicate  issues of asset selection, rebalance frequency, construction of the factor models and obtention of $X$, high-dimensional parameter estimation and updating, risk control, etc. that the problem requires would be impossible to consider here without prohibitively extending the length of the paper.

During all this section, we  therefore fix the following parameters for our model: $$N=100, \quad \mu =0,\quad X_0=\mu,$$
$A$ is diagonal with entries drawn i.i.d from a normal distribution of mean 0.5 and standard deviation  $0.1$, and the coefficients of $\sigma$ are drawn i.i.d from a uniform distribution in $[-0.3,0.3]$, except for the diagonal elements which are drawn from a uniform distribution in $[0,0.5]$. Furthermore $p=\mathds{1}$ for the first simulations, and we will also perturb it later to study different factor model regimes and the impact of imposing market neutrality. We also fix a finite horizon of $T=20$ and a temporal grid $0=t_0<t_1<\ldots t_L=T$ obtained by discretizing $[0,T]$ with constant $\Delta t=T/L=20/400=0.5$. 

From a financial perspective, the above choice of parameters means that the 100 coordinates of $X$ are correlated and mean-revert with similar speeds (given by the eigenvalues of $A$) to an equilibrium of $0$,  describing an average number of approximately 10 oscillation cycles of ups and downs before the terminal time (given by the product of $T$ and the average mean-reversion speed). The choice of $p=\mathds{1}$ arises when the asset returns themselves are mean-reverting and may be modeled directly by $X$ so we can take $\Lambda=0$ in our factor model, while the perturbations of $p$ will imply departing from this assumption to factor models with heavier loadings, in which imposing market neutrality leads to more leveraged positions. As an illustration, we plot some sample paths of the first three coordinates of such a process $X$ in Figure 1 below.
\begin{figure}[h!]
\centering
\includegraphics[width=0.75\textwidth, height=6cm]{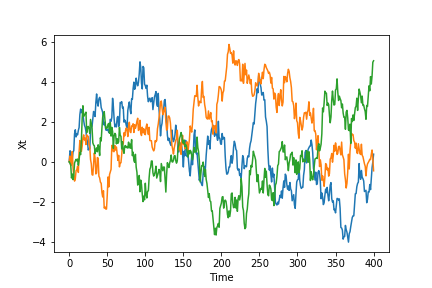}
\caption{Sample paths of the first three coordinates of $X$ in $[0,T]$}
\end{figure}

We then sample $M=1,000$ paths of $X$ on this grid exactly with standard Monte Carlo techniques by  using the fact that, since \begin{equation*}
X_{t+\Delta t}=e^{-A\Delta t}X_t+(I-e^{-A\Delta t})\mu+\int_t^{t+\Delta t}e^{-A(\Delta t+t-s)}\sigma dB_s,
\end{equation*} $X_{t+\Delta t}|X_t \sim N\left(\mu(X_t,\Delta t), \Sigma(\Delta t)\right)$, where
$$\mu(X_t, \Delta t)=e^{-A\Delta t}X_t+(I-e^{-A\Delta t})\mu, \quad\Sigma(\Delta t)=\int_0^{\Delta t} e^{-A(\Delta t-s)}\sigma\sigma' e^{-A'(\Delta t-s)}ds,$$
 and execute the following strategies\footnote{We have just simulated some simple cases of the previously discussed strategies for space limitation reasons, but it would be interesting as well to execute the strategies with some perturbations of the real parameters to simulate possible microstructural noise and imperfect estimation.} at the corresponding times $t_l$'s, with $W_0=\pi_0=0$ and $\pi_t$ constant between consecutive times:
\begin{enumerate}
\item The exponential utility strategy of Theorem 3.1 with $\gamma(t)=1,2,3,4$ and $r=0,2\%$.
\item The mean-variance utility strategy with dollar-neutrality penalty of Corollary 4.1.1 with $\gamma(t)=1,2,3,4$, $\alpha(t)=0,20,50$ and $r=2\%$.
\item The mean-variance utility strategy with quadratic transaction costs of Theorem 4.1 with $\gamma(t)=1,2,3,4$, $r=2\%$, and $C=\lambda\sigma\sigma'$ for $\lambda=0.1,0.5,1$. 
\newcounter{enumTemp}
    \setcounter{enumTemp}{\theenumi}
\end{enumerate}

Moreover, to study the result of imposing market-neutrality through the market-neutral portfolios constructed in section 2 under different factor model regimes, we perform the following additional simulation in which we experiment with the parameter $p$, which encapsulates all the factor model information and which we perturb to simulate the effect of going away from the case where the returns themselves are mean-reverting (which corresponds to the previous case $p=\mathds{1}$) and of having progressively more leveraged market-neutral portfolios:

\begin{enumerate} 
\setcounter{enumi}{\theenumTemp}
\item The three strategies above with $\gamma(t)=1,\alpha(t)=0$ and $r=2\%$ (and $\lambda = 1$ for the third strategy) for $p=\mathds{1}+\epsilon_a$ and $a=1,2,4,8$, where $\epsilon_a$ is a $N$-dimensional vector whose components are drawn i.i.d. from a uniform distribution in $[-a,a]$.
\end{enumerate}

We present the simulated  path of a sample wealth process $(W_t)_{t\in[0,T]}$, the simulated path of the first coordinate of a sample allocation process $(\pi_t)_{t\in[0,T]}$, and the histogram for the terminal wealth $W_T$ for each of the above cases in the following four subsections, along with a final analysis:

\newpage
\subsection{Simulations of the exponential-utility strategy}
  \begin{figure}[ht!]
    \centering
    \begin{subfigure}[b]{0.4\textwidth}
    \includegraphics[scale=0.4]{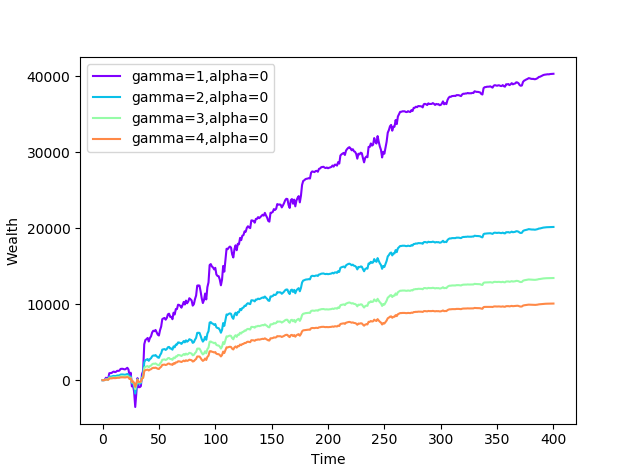}
    \caption{Sample paths of $W_t$ when $r=0$ for $\gamma=1,2,3,4$}
    \end{subfigure}\quad\quad
        \begin{subfigure}[b]{0.4\textwidth}
    \includegraphics[scale=0.4]{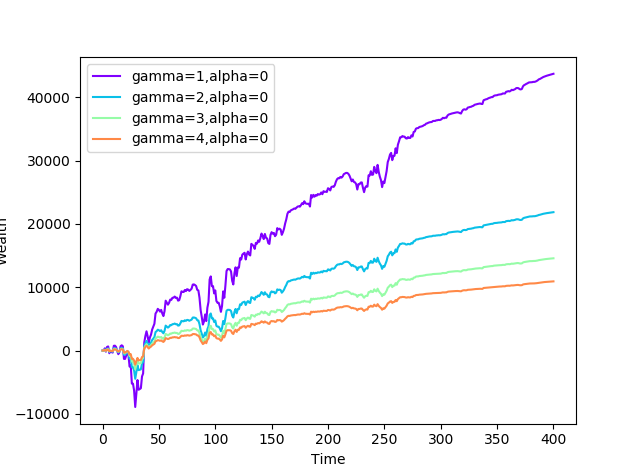}
    \caption{Sample paths of $W_t$ when $r=0.02$ for $\gamma=1,2,3,4$}
    \end{subfigure}
    \begin{subfigure}[b]{0.4\textwidth}
    \includegraphics[scale=0.4]{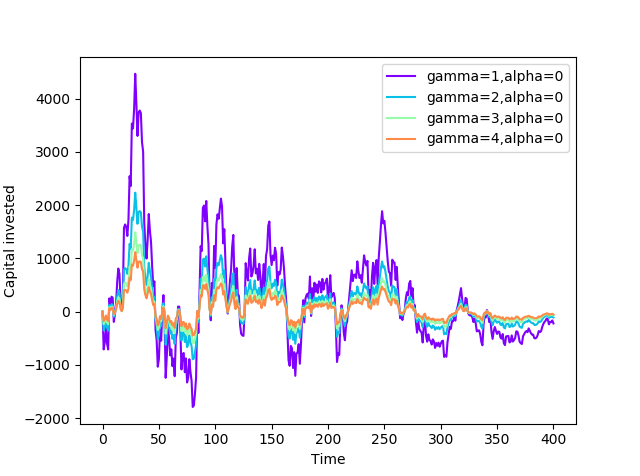}
    \caption{Sample paths of $\pi_{1t}$ when $r=0$ for $\gamma=1,2,3,4$}
    \end{subfigure}\quad\quad
        \begin{subfigure}[b]{0.4\textwidth}
    \includegraphics[scale=0.4]{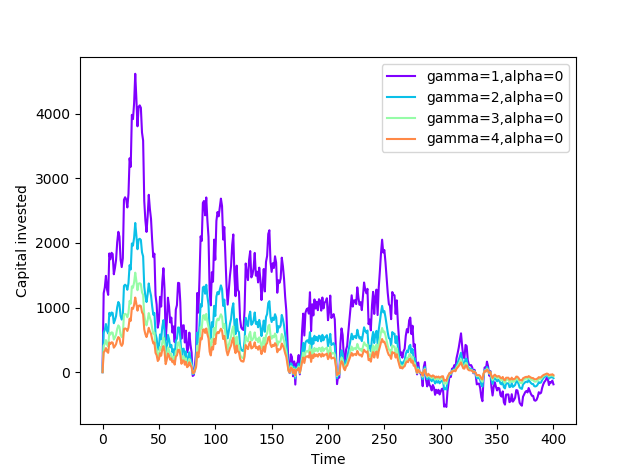}
    \caption{Sample paths of $\pi_{1t}$ when $r=0.02$ for $\gamma=1,2,3,4$}
    \end{subfigure}
        \begin{subfigure}[b]{0.4\textwidth}
    \includegraphics[scale=0.4]{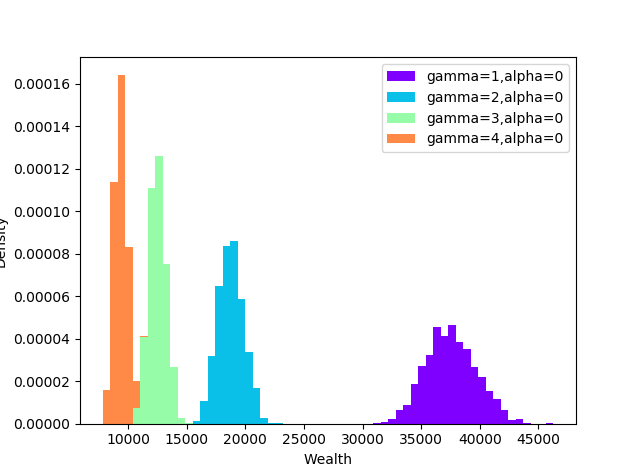}
    \caption{Histograms of $W_{T}$ when $r=0$ for $\gamma=1,2,3,4$}
    \end{subfigure}\quad\quad
        \begin{subfigure}[b]{0.4\textwidth}
    \includegraphics[scale=0.4]{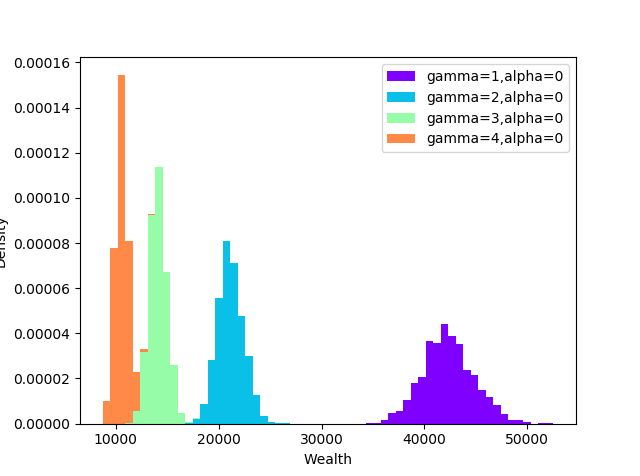}
    \subcaption{Histograms of $W_{T}$ when $r=0.02$ for $\gamma=1,2,3,4$}
    \end{subfigure}
    \caption{Results for the exponential utility}
    \end{figure}
    \clearpage
    
\subsection{Simulations of the mean-variance strategy with dollar neutrality}

      \begin{figure}[ht!!!]
    \centering
    \begin{subfigure}[b]{0.31\textwidth}
    \includegraphics[scale=0.35]{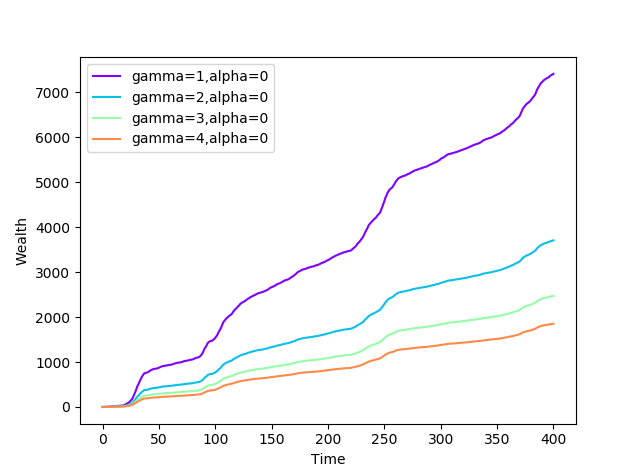}
    \caption{Sample paths of $W_t$ for $\gamma=1,2,3,4$ and $\alpha=0$}
    \end{subfigure}\quad
        \begin{subfigure}[b]{0.31\textwidth}
    \includegraphics[scale=0.35]{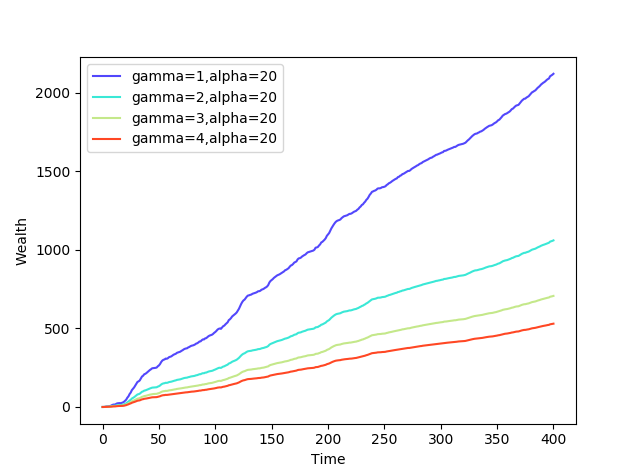}
    \caption{Sample paths of $W_t$ for $\gamma=1,2,3,4$ and $\alpha=20$}
    \end{subfigure}\quad
    \begin{subfigure}[b]{0.31\textwidth}
    \includegraphics[scale=0.35]{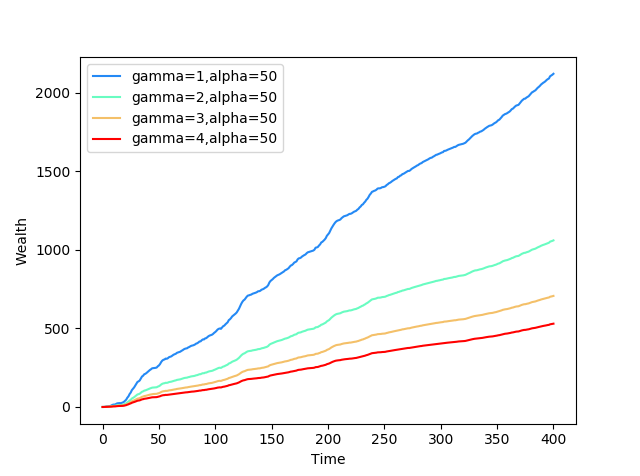}
    \caption{Sample paths of $W_t$ for $\gamma=1,2,3,4$ and $\alpha=50$}
    \end{subfigure}
       \begin{subfigure}[b]{0.31\textwidth}
    \includegraphics[scale=0.35]{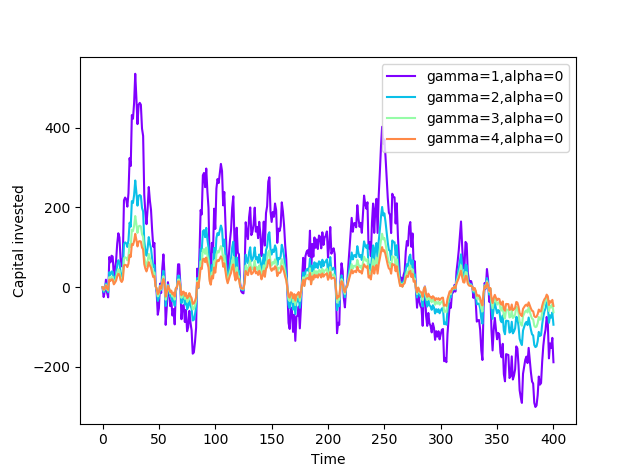}
    \caption{Sample paths of $\pi_{1t}$ for $\gamma=1,2,3,4$ and $\alpha=0$}
    \end{subfigure}\quad
        \begin{subfigure}[b]{0.31\textwidth}
    \includegraphics[scale=0.35]{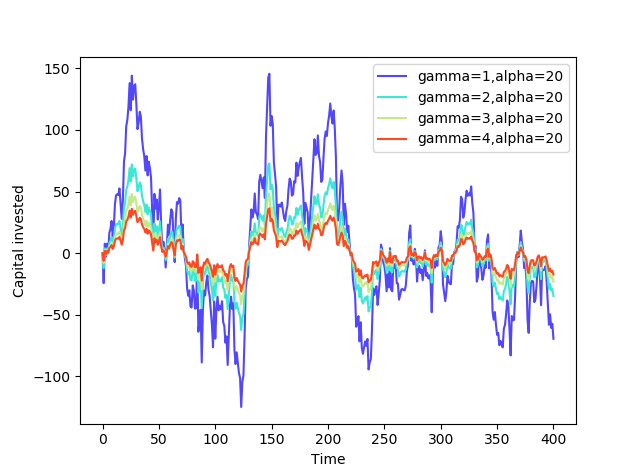}
    \caption{Sample paths of $\pi_{1t}$ for $\gamma=1,2,3,4$ and $\alpha=20$}
    \end{subfigure}\quad
    \begin{subfigure}[b]{0.31\textwidth}
    \includegraphics[scale=0.35]{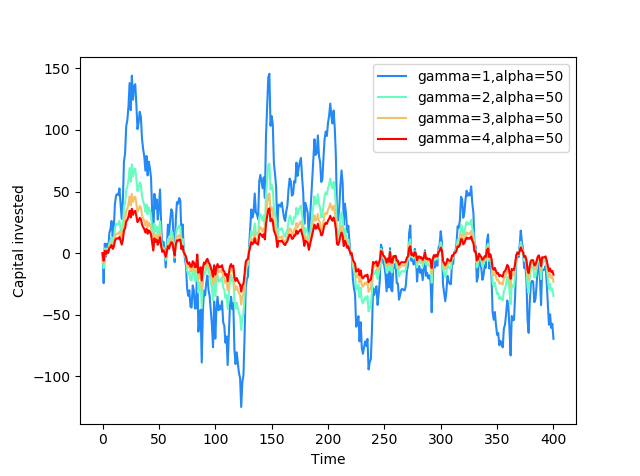}
    \caption{Sample paths of $\pi_{1t}$ for $\gamma=1,2,3,4$ and $\alpha=50$}
    \end{subfigure}
       \begin{subfigure}[b]{0.31\textwidth}
    \includegraphics[scale=0.35]{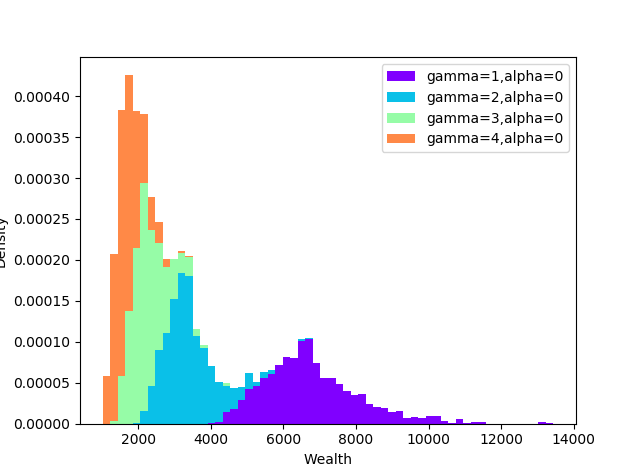}
    \caption{Histograms of $W_{T}$ for $\gamma=1,2,3,4$ and $\alpha=0$}
    \end{subfigure}\quad
        \begin{subfigure}[b]{0.31\textwidth}
    \includegraphics[scale=0.35]{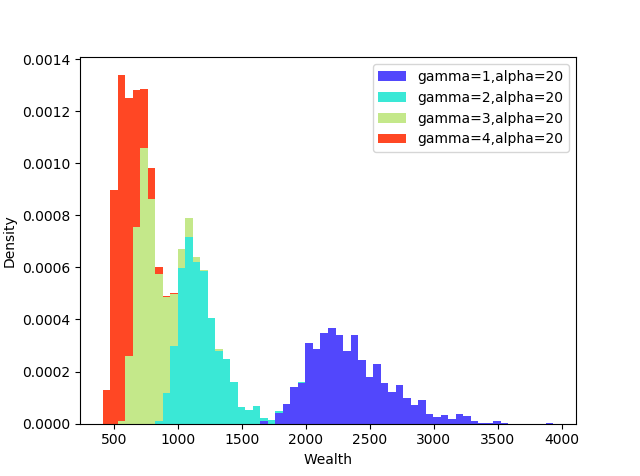}
    \caption{Histograms of $W_{T}$ for $\gamma=1,2,3,4$ and $\alpha=20$}
    \end{subfigure}\quad
    \begin{subfigure}[b]{0.31\textwidth}
    \includegraphics[scale=0.35]{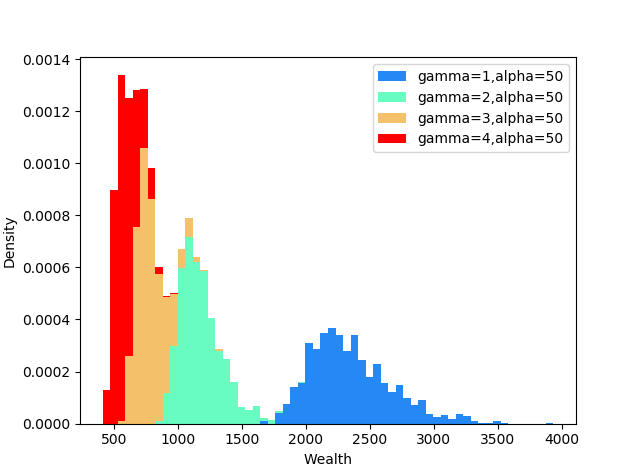}
    \caption{Histograms of $W_{T}$ for $\gamma=1,2,3,4$ and $\alpha=50$}
    \end{subfigure}
    \caption{Results for the  mean-variance utility when $r=0.02$ with different dollar-neutrality restrictions}
    \end{figure}
    \pagebreak

\subsection{Simulations of the mean-variance strategy with quadratic transaction costs}

    \begin{figure}[ht!]
    \centering
    \begin{subfigure}[b]{0.31\textwidth}
    \includegraphics[scale=0.35]{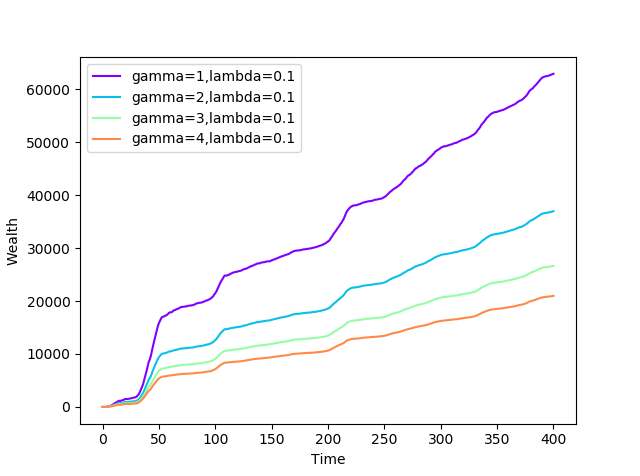}
    \caption{Sample paths of $W_t$ for $\gamma=1,2,3,4$ and $\lambda=0.1$}
    \end{subfigure}\quad
        \begin{subfigure}[b]{0.31\textwidth}
    \includegraphics[scale=0.35]{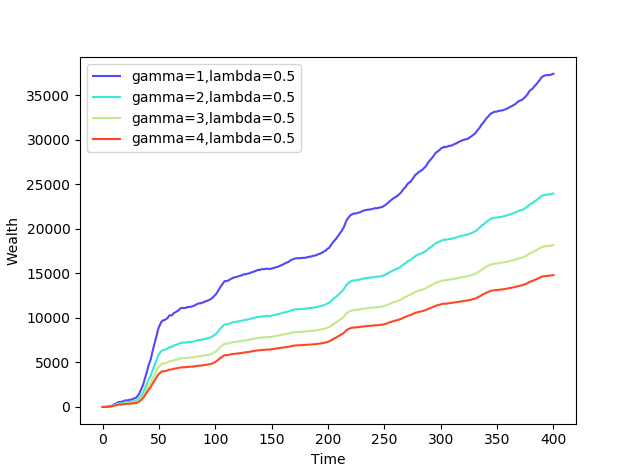}
    \caption{Sample paths of $W_t$ for $\gamma=0,1,2,3$ and $\lambda=0.5$}
    \end{subfigure}\quad
    \begin{subfigure}[b]{0.31\textwidth}
    \includegraphics[scale=0.35]{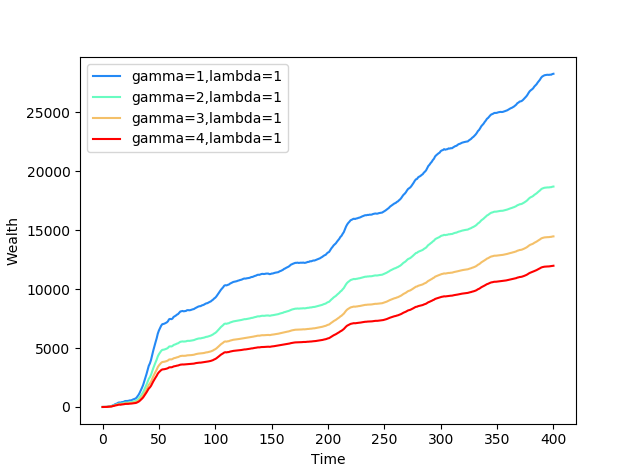}
    \caption{Sample paths of $W_t$ for $\gamma=1,2,3,4$ and $\lambda=1$}
    \end{subfigure}
       \begin{subfigure}[b]{0.31\textwidth}
    \includegraphics[scale=0.35]{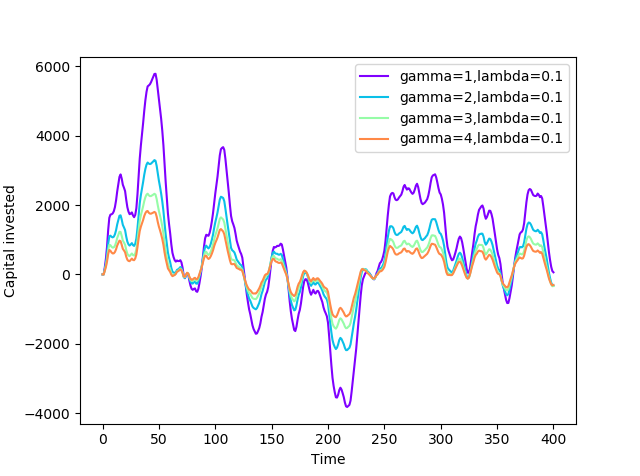}
    \caption{Sample paths of $\pi_{1t}$ for $\gamma=1,2,3,4$ and $\lambda=0.1$}
    \end{subfigure}\quad
        \begin{subfigure}[b]{0.31\textwidth}
    \includegraphics[scale=0.35]{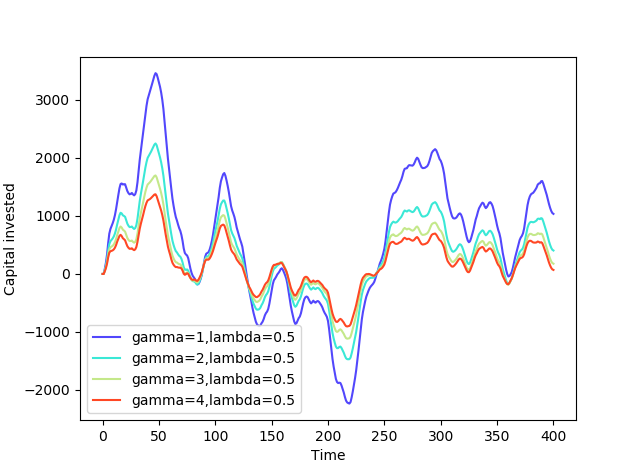}
    \caption{Sample paths of $\pi_{1t}$ for $\gamma=1,2,3,4$ and $\lambda=0.5$}
    \end{subfigure}\quad
    \begin{subfigure}[b]{0.31\textwidth}
    \includegraphics[scale=0.35]{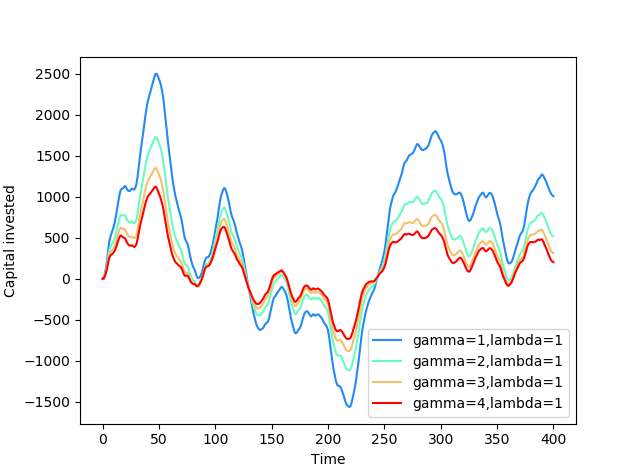}
    \caption{Sample paths of $\pi_{1t}$ for $\gamma=1,2,3,4$ and $\lambda=1$}
    \end{subfigure}
       \begin{subfigure}[b]{0.31\textwidth}
    \includegraphics[scale=0.35]{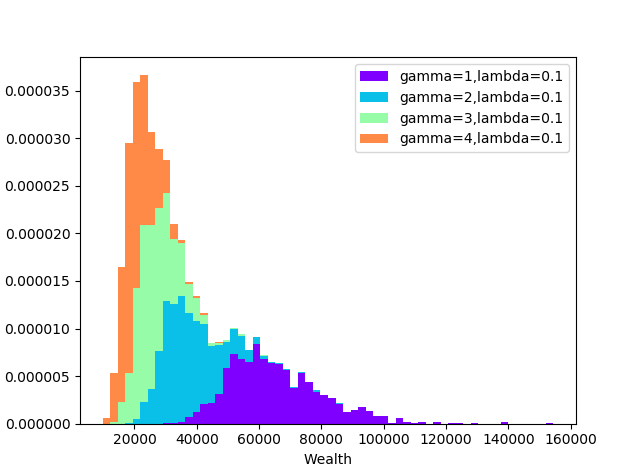}
    \caption{Histograms of $W_{T}$ for $\gamma=1,2,3,4$ and $\lambda=0.1$}
    \end{subfigure}\quad
        \begin{subfigure}[b]{0.31\textwidth}
    \includegraphics[scale=0.35]{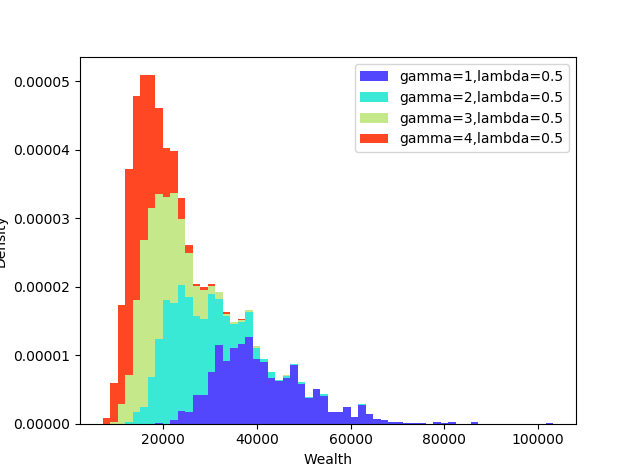}
    \caption{Histograms of $W_{T}$ for $\gamma=1,2,3,4$ and $\lambda=0.5$}
    \end{subfigure}\quad
    \begin{subfigure}[b]{0.31\textwidth}
    \includegraphics[scale=0.35]{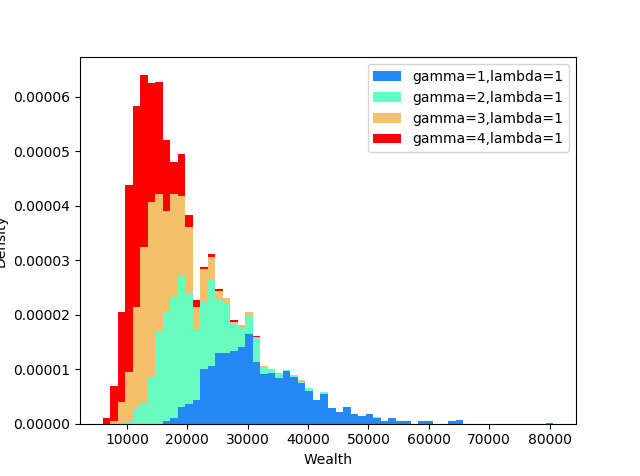}
    \caption{Histograms of $W_{T}$ for $\gamma=1,2,3,4$ and $\lambda=1$}
    \end{subfigure}
    \caption{Results for the  mean-variance utility when $r=0.02$  with different quadratic transaction costs $C=\lambda\sigma\sigma'$}
    \end{figure}
    \clearpage

\subsection{Simulations for different factor model and market-neutrality regimes}
    \begin{figure}[ht!]
    \centering
    \begin{subfigure}[b]{0.31\textwidth}
    \includegraphics[scale=0.35]{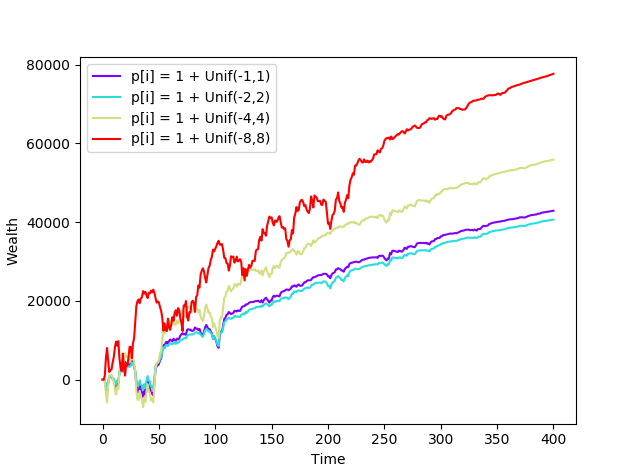}
    \caption{Sample paths of $W_t$ for the exponential utility}
    \end{subfigure}\quad
        \begin{subfigure}[b]{0.31\textwidth}
    \includegraphics[scale=0.35]{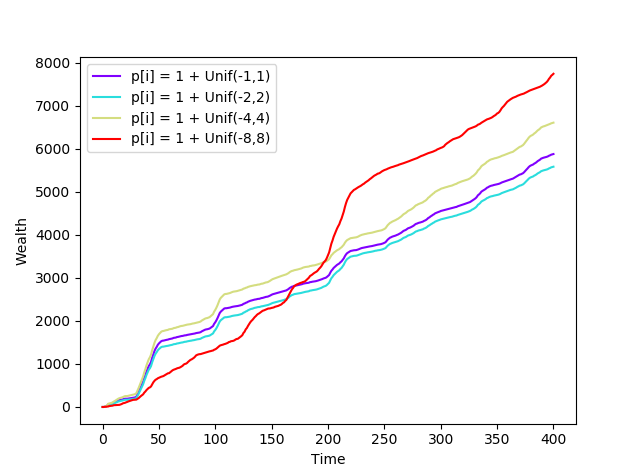}
    \caption{Sample paths of $W_t$ for the mean-variance utility}
    \end{subfigure}\quad
    \begin{subfigure}[b]{0.31\textwidth}
    \includegraphics[scale=0.35]{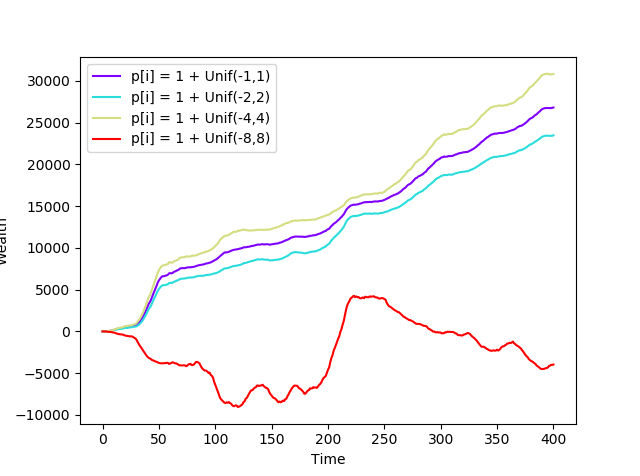}
    \caption{Sample paths of $W_t$ for the mean-variance utility with transaction costs}
    \end{subfigure}
       \begin{subfigure}[b]{0.31\textwidth}
    \includegraphics[scale=0.35]{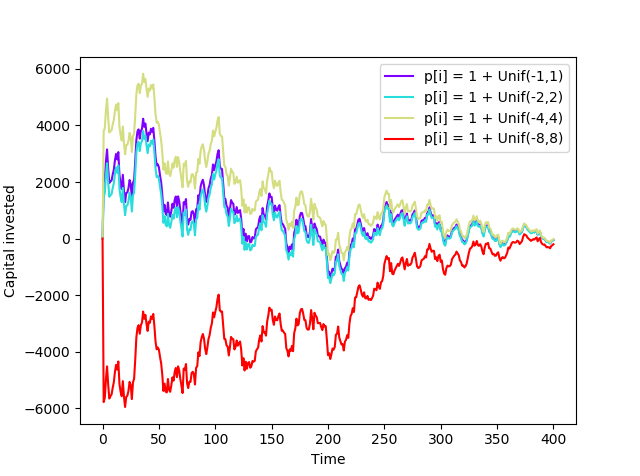}
    \caption{Sample paths of $\pi_{1t}$ for the exponential utility}
    \end{subfigure}\quad
        \begin{subfigure}[b]{0.31\textwidth}
    \includegraphics[scale=0.35]{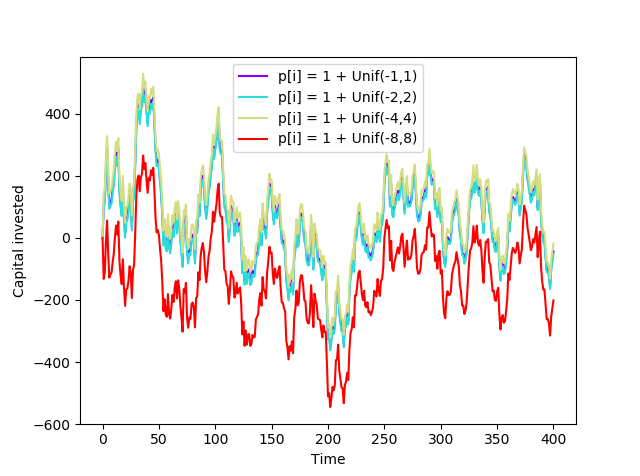}
    \caption{Sample paths of $\pi_{1t}$ for the mean-variance utility }
    \end{subfigure}\quad
    \begin{subfigure}[b]{0.31\textwidth}
    \includegraphics[scale=0.35]{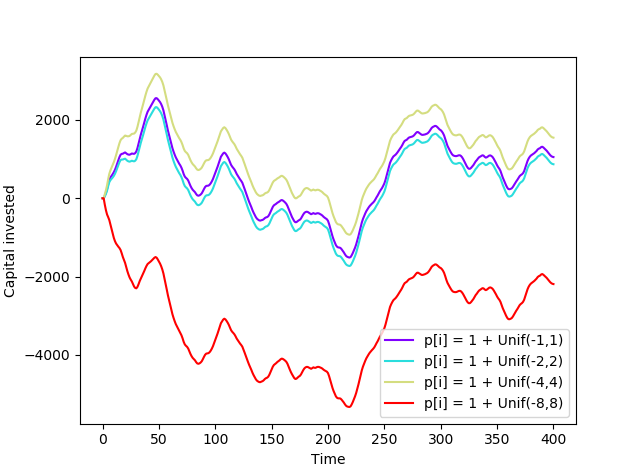}
    \caption{Sample paths of $\pi_{1t}$ for the  mean-variance utility with transaction costs}
    \end{subfigure}
       \begin{subfigure}[b]{0.31\textwidth}
    \includegraphics[scale=0.35]{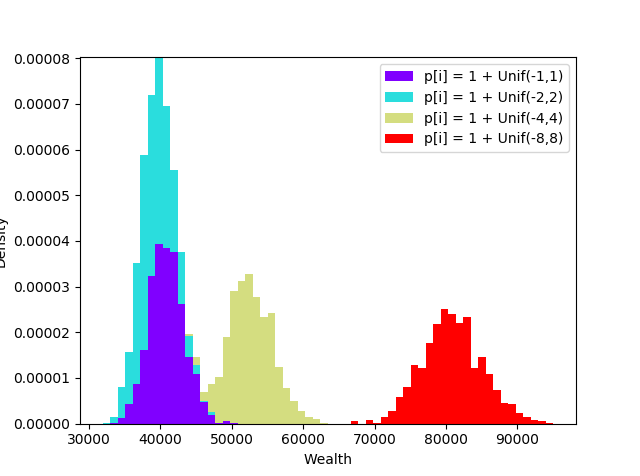}
    \caption{Histograms of $W_{T}$ for the exponential utility}
    \end{subfigure}\quad
        \begin{subfigure}[b]{0.31\textwidth}
    \includegraphics[scale=0.35]{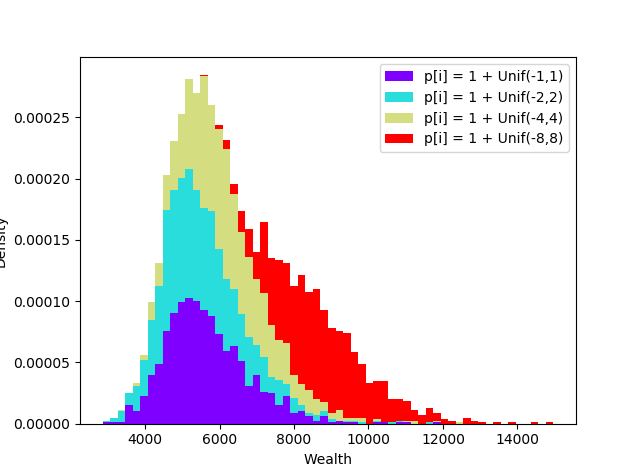}
    \caption{Histograms of $W_{T}$ for the mean-variance utility}
    \end{subfigure}\quad
    \begin{subfigure}[b]{0.31\textwidth}
    \includegraphics[scale=0.35]{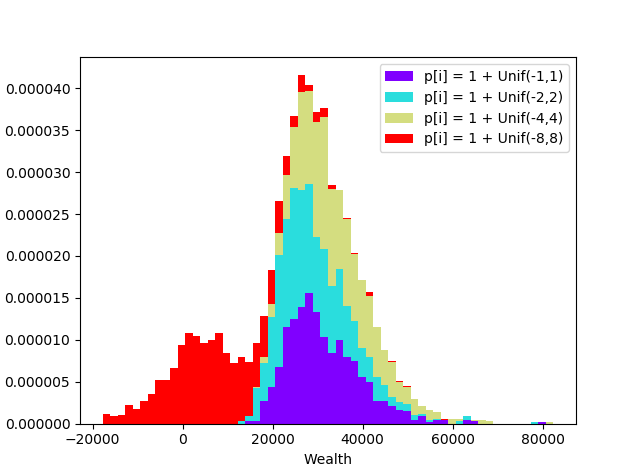}
    \caption{Histograms of $W_{T}$ for the mean-variance utility with transaction costs}
    \end{subfigure}
    \caption{Results for the three strategies above with different $p$'s, when $r=0.02, \alpha=0, \gamma=1, \lambda=1$}
    \end{figure}

\subsection{Comparison of the simulated strategies}
We now present our main conclusions after observing the previous plots, analyzing the behavior of the histograms of the final wealth $W_T$, the sample paths of the wealth process $(W_t)_{t\in[0,T]}$, and the sample paths of the positions $(\pi_{1t})_{t\in[0,T]}$, with a final subsubsection discussing the effect of imposing market-neutrality under different $p$'s.

\subsubsection{Histograms of the final wealth}
Looking first at the above histograms (Figures 3-4 (g)-(i), and 2 (e)-(f)), we see that, for our parametric choice and our setting in which $X$ is effectively a multidimensional Ornstein-Uhlenbeck process with known parameters,
\begin{enumerate}
    \item The most profitable strategy is the one derived from the exponential utility (Figure 2, (e) and (f)) with the lowest risk-aversion parameter $\gamma$, even in the most adverse scenarios of the histogram and both with and without zero interest rates. Moreover, even for bigger values of $\gamma$ this strategy significantly performs better under any regime of $r$ and $\alpha$ than the mean-variance strategy (Figure 3, (g)-(i)).
    \item We observe the following outcomes when changing one of the parameters for each of the three strategies (Figures 3-4 (g)-(i), and 2 (e)-(f)):
    \begin{itemize}
        \item Increasing the value of the risk-aversion parameter $\gamma$ produces a concentration of the density of $W_T$ around smaller values, i.e., the expected wealth decreases and so does the dispersion around it.
        \item Increasing the dollar-neutrality penalty $\alpha$ has this same negative effect, but makes little difference unless the increments in $\alpha$ are considerable.
        \item Increasing the value of the interest-rate $r$ has an overall positive effect, which is more pronounced in the mean-variance case since, as we mentioned at the end of section 2, the investor is then more aggressive than the exponential agent.
        \item Increasing the transaction cost parameter $\lambda$ decreases the expected terminal wealth, but it also skews its distribution producing a considerable right-tail (whereas all the other distributions are essentially symmetric).
    \end{itemize}
   These outcomes have a natural interpretation: since the model is perfectly specified and the parameters are known, the derived strategies will always produce benefits by construction, and they will be bigger the fewer additional constraints we impose (such as risk-aversion, dollar-neutrality, and transaction costs) and the more we can take advantage of previous success (by increasing $r$). This situation, however, might not apply under parameter misspecification, where the additional constrains would help the investor mitigate the model risk.
\end{enumerate} 

\subsubsection{Sample paths of the wealth process $W_t$}

Examining next the sample paths  for the particular simulation which is plotted (Figures 3,4 (a)-(c), and 2 (a)-(b)), we observe the same patterns as discussed in the previous paragraph when modifying the parameters $\gamma,\alpha$, $r$ and $\gamma$. There are, however, two new observations: 
\begin{enumerate}
    \item In the three strategies, after an initial period of ups and downs and similarity between the different strategies, there is a tendency towards stabilization because of the asymptotic properties of the Ornstein-Uhlenbeck process, and of differentiation depending on the parametric choices.
    \item This phenomenon is especially pronounced with the exponential utility and with bigger values of $r$ (Figure 2 (a)-(b)) since it takes more aggressive positions, reflecting the fact that sometimes the agent will invest more capital than what she will make at that moment (and sometimes even having temporary negative wealth and borrowing aggressively) to continue executing the strategy.
\end{enumerate} 

\subsubsection{Sample paths of the positions $\pi_{1t}$}
Considering now the plots of the sample paths of the positions $\pi_{1t}$ (Figures 3,4 (d)-(f), and 2 (c)-(d)), we similarly notice that
\begin{enumerate}
    \item The positions become more extreme when decreasing $\gamma$, $\alpha$ and $\lambda$ (i.e., the risk-aversion parameter, the non-dollar-neutrality penalty and the transaction cost parameter) and when increasing $r$ (the interest rate). The greatest overall impact is produced by $\gamma$ and $\lambda$ and then $r$, especially in the mean-variance case for the same reasons as before.
    \item The exponential utility strategy takes more extreme positions than the mean-variance strategies, which in this idealized setting of perfect estimation partially explains why the exponential agent obtains a greater wealth at the terminal time.
    \item The cycles in the positions $\pi_{1t}$ match the oscillations of $X_{1t}$ depicted in Figure 1, as described theoretically in the corresponding equations.
\end{enumerate}  

\subsubsection{Effects of imposing market neutrality}
Finally, looking separately at the effect of imposing market neutrality under various factor model regimes depending on $p$ (which, as we mentioned, depends quadratically on the factor model loadings), we observe the following (Figure 5):
\begin{enumerate}
    \item As the parameter $p$ gets bigger, the market neutral portfolios of section 2 become more extreme and the adopted positions $\pi_t$ also become more aggressive, especially in the exponential utility case (Figure 5, (d)-(f)).
    \item Since the strategy is more aggressive but we have perfect estimation, with bigger $p$ the mean-variance and especially the exponential strategy become more profitable. However, the wealth process also has more ups and downs (Figure 5, (a)-(c)), the standard deviation of the terminal wealth increases considerably (Figure 5, (g)-(h)), especially in the mean-variance case, and with the biggest $p$ there are also heavy losses when transaction costs are incorporated (Figure 5, (c),(i)). The strategies are therefore riskier, but a relatively large value of $p$ is needed to appreciate its effect.
    \item Lastly, note that the influence of $p$ on the strategies also depends most of the time on the value of $r$, since they normally appear combined as a factor of $rp$ in the equations describing the strategies. In particular, when $r=0$ there is no theoretical effect associated to $p$ (apart from possible model risk and high leverage in a real-world setting)  unless the dollar-neutrality parameter $\alpha(t)\neq 0$. 
\end{enumerate}

\section{Conclusions and further research}

In this paper we have aimed to provide a systematic study of high-dimensional statistical arbitrage combining stochastic control and factor models. To this end, we have first proposed a general framework based on a statistically-constructed factor model, and then shown how to obtain analytically explicit market-neutral portfolios and rephrase our problem in terms of them to make it tractable and get market neutrality by construction. Using this insight, we have studied the question of optimizing the expected  utility of the investor's terminal wealth in continuous time under both an exponential and a mean-variance objective. In both cases, we have obtained explicit closed-form solutions that avoid potentially unfeasible high-dimensional numerical methods, analyzed the corresponding strategies from the perspective of statistical arbitrage and the underlying factor model, and discussed extensions involving the addition of soft constraints on the admissible portfolios (like dollar-neutrality) and the presence of temporary quadratic transaction costs. Finally, we have run Monte Carlo simulations to explore the behavior of the previous strategies, and analyzed their qualitative aspects and their sensitivity to the relevant parameters and the underlying factor model.  

There are four natural extensions to our work, on which we are conducting research at the moment and which we intend to publish in separate papers. First, one could investigate a more realistic version of the problem in which, rather than in continuous time, the investor may only trade more realistically at an increasing sequence of optimally chosen stopping times,generalizing the literature initiated by \cite{leung-stoploss} and developing robust and efficient numerical methods. Second, it would be interesting to study more realistic modelizations of market frictions, illiquidity, and transaction costs, or to develop a model considering issues of parameter misspecification. Third, on a more empirical side and as we mentioned at the start of the section 5, one should consider in this setting the problems of construction of the factor models, high-dimensional parameter estimation, and risk control,  along with out-of-sample experiments with real market data under the strategies developed in this paper. Fourth and finally, one could study a more data-driven version of the problem, where the fixed stochastic model is replaced by new tools from reinforcement learning.

\section{Acknowledgements}
The author would like to thank George Papanicolaou for suggesting the topic of the previous research and for many insightful discussions about the problem and the presentation of the results, and the editor and an anonymous reviewer for their very helpful suggestions to improve the quality of the paper.
\newpage
 
\bibliography{bib}

\newpage
\appendix
\section{Appendix: Proofs}
\subsection{Proof of Theorem 3.1}
\textbf{Proof of Proposition 3.1: }\\
The dynamic programming principle suggests that the value function $H$ should satisfy the dynamic programming equation (\ref{DPE}) with terminal condition $H(T,x,w)=-e^{-\gamma w}$. The optimal control may then be found in feedback form by looking at the first order condition of the term inside the supremum, since the corresponding function is quadratic and concave in $\pi$ (if $\partial_{ww}H<0$, i.e., if there is risk aversion). The first order condition gives that
$$0=\sigma\sigma'\partial_{ww}H\pi +\left(A(\mu - x)-pr\right)\partial_wH + \sigma\sigma'\nabla_{xw}H,   $$
and solving for $\pi$ we find the control given in the proposition's statement. Putting it back into (\ref{DPE}) we get the following non-linear and $(N+2)$-dimensional PDE
\begin{equation}\label{HJBfeedback}
0=\partial_tH + (A(\mu - x))'\nabla_xH+\frac{1}{2}\mathrm{Tr}(\sigma\sigma'\nabla_{xx}H)+wr\partial_wH\ -\frac{\mathcal{D}H'(\sigma\sigma')^{-1}\mathcal{D}H}{2\partial_{ww}H}.
\end{equation} 
Now, looking at the terminal condition, we guess that the solution of this PDE will be of the form $H(t,x,w)=-\exp(-\gamma(we^{r(T-t)}+h(t,x)))$ 
for some function $h(t,x)$ to be determined and such that $h(T,x)=0$. Some easy computations then show that 
$$\partial_tH = -\gamma H(-rwe^{r(T-t)}+\partial_t h)\hspace{0,5cm}\partial_wH = -\gamma e^{r(T-t)} H \hspace{0,5cm} \partial_{ww}H =\gamma^2e^{2r(T-t)} H \hspace{0,5cm}
\nabla_{xw}H =\gamma^2e^{r(T-t)}  H\nabla_xh$$
$$\nabla_xH =-\gamma H\nabla_xh
\hspace{0,6cm}
\nabla_{xx}H =-\gamma H(\nabla_{xx}h-\gamma \nabla_xh\nabla_xh') \hspace{0,6cm} \mathcal{D}H =-\gamma e^{r(T-t)} H (A(\mu-x)-pr-\gamma  \sigma\sigma'\nabla_xh).$$ 
Plugging all this into (\ref{HJBfeedback}), dividing everything by $-\gamma H$, and doing some simple algebra to expand the last term yields
\begin{multline*} 
0 = -rwe^{r(T-t)}+\partial_th + (A(\mu - x))'\nabla_xh+\frac{1}{2}\mathrm{Tr}(\sigma\sigma' (\nabla_{xx}h-\gamma\nabla_{x}h\nabla_{x}h'))+wre^{r(T-t)}\ +\\
\frac{1}{2\gamma}(A(\mu-x)-pr)'(\sigma\sigma')^{-1}(A(\mu-x)-pr)+\frac{\gamma}{2}\nabla_xh'\sigma\sigma'\nabla_xh - (A(\mu-x)-pr)'\nabla_xh
\end{multline*}
and we can see that the non-linear terms in $h$, the terms in $w$, and the third and part of the last term of the PDE get cancelled and the equation gets considerably simplified, obtaining 
$$0 = \partial_th + \frac{1}{2}\mathrm{Tr}(\sigma\sigma' \nabla_{xx}h)+rp'\nabla_xh+\frac{1}{2\gamma}(A(\mu-x)-pr)'(\sigma\sigma')^{-1}(A(\mu-x)-pr).$$
This is now a parabolic linear PDE in $h$ and we can find explicitly its solution by using the Feynman-Kac formula. Indeed, if we consider the stochastic process given by
\begin{equation}
\label{SDEhelper}
dY_t=rpdt+\sigma dB^*_t
\end{equation}
we can rewrite the above equation in terms of the infinitesimal generator $\mathcal{L}^*$ of $Y$ as
$$0=(\partial_t+\mathcal{L}^*)h+\frac{1}{2\gamma}(A(\mu-x)-pr)'(\sigma\sigma')^{-1}(A(\mu-x)-pr)$$
and then we can express its solution via the following conditional expectation, which is the probabilistic representation given in the proposition's statement:
\begin{align*}
h(t,x)&=\E_{t,x}^*\left[\int_t^T \frac{1}{2\gamma}(A(\mu-Y_s)-pr)'(\sigma\sigma')^{-1}(A(\mu-Y_s)-pr)\ ds\right]\\
&=\frac{1}{2\gamma}(A\mu-pr)'(\sigma\sigma')^{-1}(A\mu-pr)(T-t)\\
&\hspace{0,4cm} -\frac{1}{\gamma}(A\mu-pr)'(\sigma\sigma')^{-1}A\E_{t,x}^*\left[\int_t^TY_sds\right]+\frac{1}{2\gamma}\E_{t,x}^*\left[\int_t^TY_s'A'(\sigma\sigma')^{-1}AY_sds\right].
\end{align*} 
Finally, to find $h$ explicitly, notice that we can easily solve the SDE (\ref{SDEhelper}), obtaining, for $s\geq t$,
$$Y_s = x +rp(s-t)+\sigma(B^*_s-B^*_t).$$
and this allows us to compute the two expectations in our expression for $h$ above. Indeed, Fubini's theorem and elementary facts about the Brownian motion immediately yield
$$\E_{t,x}^*\left[\int_t^TY_sds\right] = \int_t^T\E_{t,y}^*\left[Y_s\right]ds=x(T-t)+rp\frac{(T-t)^2}{2} $$

and, interchanging integral and expectation again and noticing that 
$$\E^*[(B^*_s-B^*_t)'\sigma'A'(\sigma\sigma')^{-1}A\sigma(B^*_s-B^*_t)]=(s-t)\mathrm{Tr}(\sigma'A'(\sigma\sigma')^{-1}A\sigma), $$
we similarly find out that
\begin{multline*}
\E_{t,x}^*\left[\int_t^TY_s'A'(\sigma\sigma')^{-1}AY_sds\right] =\int_t^T\left(x +rp(s-t)\right)'A'(\sigma\sigma')^{-1}A\left(y +rp(s-t)\right)+ (s-t)\mathrm{Tr}(\sigma'A'(\sigma\sigma')^{-1}A\sigma)ds  \\ 
= x'A'(\sigma\sigma')^{-1}Ax(T-t)+\left(2x'A'(\sigma\sigma')^{-1}Arp+\mathrm{Tr}(\sigma'A'(\sigma\sigma')^{-1}A\sigma)\right)\frac{(T-t)^2}{2}+r^2p'A'(\sigma\sigma')^{-1}Ap\frac{(T-t)^3}{3},
\end{multline*}
which gives us the complete explicit solution of the DPE, and hence the explicit form of the optimal strategy $\pi^*$ by using equation (\ref{optimalpi}). \qed \\  

\textbf{Proof of Proposition 3.2:}\\
Since in the previous proof we have found explicitly the classical smooth solution $H$ of the dynamic programming equation, we just have to check that $\pi^*\in\mathcal{A}_{[0,T]}$ and that the usual regularity conditions hold for the classical proof to apply. More precisely, this means that the local martingale $dH-\mathcal{L}_{t,x,w}^\pi Hdt$ is a supermartingale for any admissible $\pi$ and a true martingale for $\pi^*$, where $\mathcal{L}_{t,x,w}^\pi$ is the infinitesimal generator of the controlled process $(X,W^\pi)$, or some sufficient condition for this like the one we stated in Proposition 3.2 in terms of the model parameters, which is what we will show here.

As for the first issue, it is easy to see that $\pi^*\in\mathcal{A}_{[0,T]}$. Indeed, it is obviously $\mathcal{F}_t$-adapted and predictable (in fact, it has continuous paths) and, using the trivial inequalities $||x+y||^2\leq 2||x||^2+2||y||^2$ and $||Ax||\leq ||A||||x||$ and the fact that $X_t$ is a Gaussian process, it is easy to see that $\int_0^T\E[||\pi_s^*||^2]ds<\infty$. 
Moreover, applying Ito's formula to the process $e^{-rt}W_t$ yields
$$d(e^{-rt}W_t)=-re^{rt}W_tdt+e^{-rt}dW_t=\pi_t\cdot e^{-rt}dX_t-\pi_t\cdot e^{-rt}prdt$$
and, therefore,
\begin{equation}\label{optimalW}
W_t=w+e^{rt}\left(\int_0^t \pi_s\cdot e^{-rs}dX_s-\int_0^t\pi_s\cdot e^{-rs}prds\right)
\end{equation}
for any $t\geq 0$ and any admissible control $\pi$. Thus, the SDE for $W$ has a unique strong solution $W^*$ for the particular case $\pi = \pi^*$ for any initial data, given by the above integral (note that the stochastic integral is well defined, since $dX_s=A(\mu-X_s)ds+\sigma dB_s$, $\pi^*$ and $X$ are continuous, and again $\int_0^Te^{-2rs}\E[||\pi_s'^*\sigma||^2]ds<\infty$).

As for the regularity conditions, we adapt the proof of Theorem 2.1. of \cite{Tourin2016}, which guarantee the uniform $\Pp$-integrability of the family of random variables $(H(\tau,X_\tau,W_\tau^*))_{\tau\in[0,T]}$ where $\tau$ is a $\mathcal{F}$-stopping time, and which we adapt to the parameters of the present model obtaining the sufficient conditions stated in Proposition 3.2. 

The key observation to adapt their proof is that in our case we also have that the hypothesized value function is of the form $H(t,x,w)=-\exp(-\gamma we^{r(T-t)}-\frac{1}{2}x'A_2(t)x-A_1(t)x-A_0(t))$ for some explicit smooth functions $A_i(t)$ that we computed in the proof of Proposition 3.1, and
our $X$ is also a matrix Ornstein-Uhlenbeck process under $\Pp$ with SDE $dX_t=A(\mu-X_t)dt+\sigma dB_t$, and
$$\gamma W^*_\tau e^{r(T-\tau)}=\gamma we^{r(T-\tau)}+\gamma\int_0^\tau \pi_s^*\cdot e^{r(T-s)}(A(\mu-X_s)-pr)ds+\gamma\int_0^\tau \pi_s^*\cdot e^{r(T-s)}\sigma dB_s$$
as we showed in (\ref{optimalW}). Thus, using the Cauchy-Schwarz inequality as in their proof, the part corresponding to $-\frac{1}{2}X_\tau'A_2(\tau)X_\tau-A_1(\tau)X_\tau-A_0(\tau)$ in the above expression for $H(\tau,X_\tau,W^*_\tau)$ may be bounded following their reasoning. As for the part corresponding to $-\gamma W_\tau^* e^{r(T-\tau)}$, we can again repeat their argument, but noting that the quadratic term in $X_s$ in the first integral above is now
$X_s'C_0(s)X_s$ for the matrix $C_0(s)$ that we defined before, and likewise the term in $X_s$ in the second integral is $X_s'C_1(s)$, which following their proof gives respectively the two explicit sufficient conditions that we stated in Proposition 3.2.  \qed

\subsection{Proof of Theorem 3.2}
The proof of this follows the same lines as the previous one and is simpler, so we just indicate the relevant changes. The HJB equation is now
\begin{multline*}
0=\partial_tH + (\mu - x)'A'\nabla_xH+\frac{1}{2}\mathrm{Tr}(\sigma\sigma'\nabla_{xx}H)\ + \\
\sup_\pi\left(\left(\pi'A(\mu - x)+(w-\pi'p)r\right)\partial_wH+\frac{1}{2}\pi'\sigma\sigma'\pi\partial_{ww}H + \pi'\sigma\sigma'\nabla_{xw}H -\frac{\gamma(t)}{2}\pi'\sigma\sigma'\pi \right)
\end{multline*}

with terminal condition $H(T,x,w)=w.$

Guessing that the value function will now be of the form $H(t,x,w)=we^{r(T-t)}+a(t)+b(t)'x+\frac{1}{2}x'c(t)x$ for a scalar $a(t)$, an $N$-dimensional vector $b(t)$, and a symmetric $N\times N$ matrix $c(t)$, and plugging this into the above equation, we obtain the hypothesized optimal control given in the statement of the theorem and the above PDE gets reduced to the following system of three first-order linear matrix ODEs
$$0=\partial_tc-A'c-cA+e^{2r(T-t)}A'(\gamma(t)\sigma\sigma')^{-1}A $$
$$ 0 = \partial_tb-A'b+cA\mu-e^{2r(T-t)}A'(\gamma(t)\sigma\sigma')^{-1}(A\mu-pr)$$
$$0 = \partial_ta+\frac{1}{2}\left(\mu'A'b+b'A\mu\right)+\frac{1}{2}\mathrm{Tr}(\sigma\sigma'c)+\\
\frac{e^{2r(T-t)}}{2}(A\mu-pr)'(\gamma(t)\sigma\sigma')^{-1}(A\mu-pr)$$
with terminal conditions $a(T)=b(T)=c(T)=0$.

This system has an explicit bounded solution in $[0,T]$, since the classical solution of the general first-order linear matrix ODE $\partial_ty+uy+v(t)=0$ with $y(T)=0$ is given by 
$$y(t)=\int_t^T\exp\left((s-t)u\right)v(s)ds,$$
if $v(s)$ is continuous on $[0,T]$, in which case it is automatically bounded on $[0,T]$ as well; and similarly the classical solution of $\partial_ty+uy+yu'+v(t)=0$ with $y(T)=0$ for a symmetric $v$ is given by
$$y(t)=\int_t^T\exp\left((s-t)u\right)v(s)\exp\left((s-t)u'\right)ds.$$

Thus, the HJB equation has an explicit classical solution which has \textit{quadratic growth in the state variables uniformly in} $t$. A classical verification result (cf. for example Theorem 4.3 of \cite{Guyon}) then yields that our hypothesized optimal control is indeed optimal provided that it is admissible, which may be checked exactly as in the proof of Theorem 3.1. \qed

\subsection{Proof of Theorem 4.1}
\textbf{Proof of Proposition 4.1:}\\ 
The corresponding dynamic programming equation is in this case
\begin{multline*}
0=(\partial_t + \mathcal{L}_X)H+\left(\pi'A(\mu - x)+(w-\pi'p)r\right)\partial_wH+\frac{1}{2}\pi'\sigma\sigma'\pi\partial_{ww}H+ \\
 + \pi'\sigma\sigma'\nabla_{xw}H -\frac{\gamma(t)}{2}\pi'\sigma\sigma'\pi +\sup_I\left(I'\nabla_\pi H-\frac{1}{2}I'CI\right)
\end{multline*}
with terminal condition $H(T,x,w,\pi)=w$ and where the supremum is obviously attained at $I^*=C^{-1}\nabla_\pi H$.

Substituting this back in the above equation and plugging the stated ansatz we obtain that
\begin{multline*}
0=\frac{1}{2}\pi'\partial_t a\pi+\pi'(\partial_t+\mathcal{L}_X)b+(\partial_t+\mathcal{L}_X)d+\pi'(A(\mu-x)-pr)e^{r(T-t)} \\
 -\frac{\gamma(t)}{2}\pi'\sigma\sigma'\pi+\frac{1}{2}(a\pi+b)'C^{-1}(a\pi+b).
\end{multline*}
Matching the coefficients for $\pi'(\cdot)\pi$, $\pi'(\cdot)$, and the constant yields the above differential equations.  \qed

Before we prove the next proposition, we state here the following result for comparison and existence of solutions of matrix Riccati ODEs (cf. Theorem 2.2.2 in \cite{ComparisonRiccati}), which we will use in our proof. \\

\begin{theorem}\label{ExistenceRiccati}
Let $L_1(t), L_2(t), L(t), N_1(t), N_2(t) \in \R^{d\times d}$ be piecewise continuous on $\R$. Moreover, suppose $L_1(t), L_2(t), N_1(t), N_2(t)$ and $S_1, S_2 \in \R^{d\times d}$ are symmetric. Let $T > 0$ and 
$$S_1 \geq S_2, L_1\geq  L_2\geq  0, N_1\geq  N_2 ,$$
on $[0, T]$. Assume that the terminal value problem
$$\partial_tH_1 +H_1L_1H_1 +MH_1 +H_1M+N_1 = 0 ,\quad H_1(T) = S_1 ,$$
has a (symmetric) solution $H_1$ on $[0, T]$. Then the terminal value problem
$$\partial_t H_
2 +H_2L_2H_2 +MH_2 +H_2M+N_2 = 0 ,\quad H_2(T) = S_2 ,$$
has a (symmetric) solution $H_2$ on $[0, T]$ and $H_1(t) \geq  H_2(t)$ for all $t\in [0, T].$
\end{theorem}

We are now in a position to give the following:

\textbf{Proof of Proposition 4.2:}
\begin{enumerate}
\item The first statement follows directly from the comparison Theorem \ref{ExistenceRiccati} stated before, since the matrix Riccati ODE
$$\partial_t a+aC^{-1}a=0$$
with terminal condition $a(T)=0$ has the obvious symmetric solution $a(t)=0$ defined on all $[0,T]$. Thus, (\ref{RiccatiCosts}) has a symmetric classical solution $a(t)$ defined on all $[0,T]$ with $a\leq 0$, which is bounded because $[0,T]$ is compact and $a$ is differentiable hence continuous.

As for the particular solution when $\gamma(t)=\gamma$, simply note that pre- and post-multiplying (\ref{RiccatiCosts}) by $C^{-1/2}$ and defining $\tilde{a}:=C^{-1/2}aC^{-1/2}$ gives the new Riccati
$$\partial_t \tilde{a}-\gamma C^{-1/2}\sigma\sigma'C^{-1/2}+\tilde{a}^2=0,$$
whose solution is $\tilde{a}(t)=D\tanh(D(t-T))$.

\item The existence of solutions with polynomial growth and their probabilistic representation in the above form follow from a vector-valued version of the Feynman-Kac theorem (see Appendix A.3 of \cite{Cartea2018} for a proof of how to adapt the one-dimensional case) provided that the appropriate regularity conditions hold. Using, for example, Condition 2 of Appendix E in \cite{Duffie}, it is sufficient that all the functions of $(t,x)$ $A(\mu-x)$, $\sigma$, $a(t)'C^{-1}$, $e^{r(T-t)}(A(\mu-x)-rp)$ (and $b(t,x)'C^{-1}b(t,x)$ for the existence of $d$) are uniformly Lipschitz in $x$, they and their first and second derivatives in $x$ are continuous with polynomial growth in $x$ uniformly in $t$, and $a(t)\leq 0$. All of these properties are straightforward to check in this case because all the corresponding functions are given explicitly and are simple, and the required properties for $a$ follow from (1). 

The fact that $b$ has linear growth in $x$ uniformly in $t$ is then a consequence of the probabilistic representation (\ref{Probrepb}). Indeed, Fubini's theorem implies that
$$b(t,x)=\int_t^T:\exp\left(\int_t^sa'(u)C^{-1}du\right):e^{r(T-s)}(A\E_{t,x}\left[\mu-X_s\right]-rp)ds$$
whereas the fact that
 \begin{equation*}\label{SolutionOU}
X_{t+\Delta t}=e^{-A\Delta t}X_t+(I-e^{-A\Delta t})\mu+\int_t^{t+\Delta t}e^{-A(\Delta t+t-s)}\sigma dB_s
\end{equation*}
yields
$$\E_{t,x}\left[\mu-X_s\right]=e^{-A(s-t)}(\mu-x).$$
Combining the two pieces and using the boundedness of $a$ and the compactness of $[0,T]$ gives the desired uniform bound in $t$.

The quadratic growth of $d$ in $x$ uniformly in $t$ is then obvious looking at its probabilistic representation and using the linear growth of $b$.\qed\end{enumerate}

\textbf{Proof of Proposition 4.3:}\\
Combining the two previous propositions, we have already found an explicit classical solution of the associated HJB equation with quadratic growth in the state variables uniformly in $t$, so using again Theorem 4.3 in \cite{Guyon}, we just have to verify that the candidate  intensity given in Proposition 4.1 is admissible.

For this, first of all note that the corresponding SDEs controlled by the above intensity have a unique strong solution for any initial data. Indeed, given $I^*$ and the definition of $I$ as $d\pi=Idt$, we can solve explicitly the corresponding first-order linear matrix ODE for $\pi^*$ yielding, for $s\geq t$,
$$\pi_s^*=\pi_t+\int_t^s:\exp\left(\int_u^s\mathrm{Rate}(v)dv\right):\mathrm{Aim}(u,X_u)du,$$
and this $\pi^*$ in turn defines $W^*$ like in the proof of Theorem 3.1.

Finally, from the above construction it is obvious that both $\pi^*_t$ and $I^*_t$ are $\mathcal{F}_t$-adapted and predictable (in fact, they have continuous paths), and the property that $\pi^*$ is in $L^2([0,T]\times\Omega)$ (i.e., that $\int_0^T\E[||\pi_s^*||^2]ds<\infty$) stems from the observation that $\mathrm{Rate}(u)$ is deterministic and bounded (because of Proposition 4.2.1), $\mathrm{Aim}(t,x)$ has linear growth in $x$ uniformly in $t$ (by Proposition 4.2.2), and $X$ is a Gaussian process (so it is in $L^2([0,T]\times\Omega)$). 

$I_t^*$ is likewise in $L^2([0,T]\times\Omega)$ since, as we saw in Proposition 4.1, $I_t^*=C^{-1}(a(t)\pi_t^*+b(t,X_t))$ and we can therefore use the triangular inequality, the just shown fact that $\pi_t^*$ is in $L^2([0,T]\times\Omega)$, and the same arguments as above that $a(t)$ is bounded (because of Proposition 4.2.1), that $b(t,x)$ has linear growth in $x$ uniformly in $t$ (by Proposition 4.2.2), and that $X$ is a Gaussian process, to conclude. \qed

\end{document}